\documentclass[reprint,NumberedRefs]{JASAnew}

\hypersetup{
    colorlinks=true,
    linkcolor=blue,
    filecolor=magenta,      
    urlcolor=cyan,
    pdftitle={Overleaf Example},
    pdfpagemode=FullScreen,
    }
\usepackage{xcolor}
\definecolor{shadecolor}{rgb}{0.95,0.95,0.95}
\definecolor{frameborder}{rgb}{0,0,0}

\begin{document}

\nolinenumbers

\title{Joint Spatio-Temporal Discretisation of Nonlinear Active Cochlear Models}

\author{Ting Dang}
\affiliation{Department of Computer Science and Technology, University of Cambridge, Cambridge, UK}
\affiliation{School of Electrical Engineering and Telecommunications, University of New South Wales, Sydney, NSW 2052, Australia}
\author{Vidhyasaharan Sethu}
\author{Eliathamby Ambikairajah}
\author{Julien Epps}
\affiliation{School of Electrical Engineering and Telecommunications, University of New South Wales, Sydney, NSW 2052, Australia}
\author{Haizhou Li}
\affiliation{Department of Electrical and Computer Engineering,
National University of Singapore, Singapore}


\begin{abstract}
Biologically inspired auditory models play an important role in developing effective audio representations that can be tightly integrated into speech and audio processing systems. Current computational models of the cochlea are typically expressed in terms of systems of differential equations and do not directly lend themselves for use in computational speech processing systems. Specifically, these models are spatially discrete and temporally continuous. This paper presents a jointly discretised (spatially and temporally discrete) model of the cochlea which allows for processing at fixed time intervals suited to discrete time speech and audio processing systems. The proposed model takes into account the active feedback mechanism in the cochlea, a core characteristic lacking in traditional speech processing front-ends, which endows it with significant dynamic range compression capability. This model is derived by jointly discretising an established  semi-discretised (spatially discrete and temporally continuous) cochlear model in a state space form. We then demonstrate that the proposed jointly discretised implementation matches the semi-discrete model in terms of its characteristics and finally present stability analyses of the proposed model.
\end{abstract}

\maketitle
\section{\label{sec:1} Introduction}
Computational modelling of cochlear mechanics has served as an important role in improving the understanding of the physical behaviour of the peripheral auditory system, and further lays the foundations for computational speech analyses. These models can aid in a wide range of applications including the detection of abnormal hearing losses, development of cochlear implants, and potentially serve as front-ends in speech and audio processing systems. A key element of the cochlea is the active mechanism which functions as a nonlinear amplifier and provides an increased sensitivity and frequency selectivity via suitable feedback paths within the cochlea \cite{elliott2012cochlea,camalet2000auditory}. This continuously adaptive operation of the cochlea helps ensure that the neuronal representation of the sounds is relatively invariant across a large dynamic range of input sounds. Most state-of-the-art speech processing front-ends, which use time-invariant filterbanks, lack this ability to accommodate input signals with large dynamic ranges.
\begin{figure}[b]
\colorbox{shadecolor}{\parbox{\dimexpr\linewidth-2\fboxsep}{
The following article has been submitted to  \textit{The Journal of the Acoustical Society of America.    }     After it is published,  it  will  be  found  at:     \url{http://asa.scitation.org/journal/jas}.    }}
\end{figure}
The basilar membrane (BM) is the key structural element within the cochlea which serves as a spectrum analyser, where each different position along the BM responds to stimuli of different frequencies \cite{von1960experiments,moore2012introduction}. Existing mathematical models of the cochlea involve a wave propagating along the BM, generated by an interaction between the inertia of the fluid in the chambers of the cochlea and the stiffness of BM. The BM responses can be either solved via Wentzel-Kramers-Brillouin (WKB) methods \cite{steele1974behavior,steele1980improved,taber1981cochlear} carried out directly in the continuous domain \cite{lim2002three,kanis1993self,chadwick1998compression}, or via finite element models which divide both the BM and the fluid pressure generated by the inertia of the fluid in the chambers into a number of discrete elements. The WKB approach imposes a variety of inherent assumptions and incurs relatively high computational complexity \cite{ni2014modelling}, while the finite element approach is more computationally convenient. However, the finite element approach leads to a set of ordinary differential equations (ODEs) \cite{elliott2007state} and requires numerical solution methods with adaptive step sizes, making them unsuitable to be directly used in discrete-time speech and audio processing systems which employ computations at fixed time steps.

While active cochlear models have not seen widespread use in speech processing systems, there have been multiple attempts to use passive cochlear models (models that do not include active feedback). Most commonly, passive cochlear models in speech recognition or sound recognition tasks are implemented as filters in the equivalent rectangular bandwidth (ERB) scale or as gammatone filters \cite{sharan2015cochleagram,buermannspeech}. Passive cochlear models also tend to be implemented as parallel filterbanks, which do not capture the longitudinal coupling properties of the BM. Finally, a handful of studies have attempted to use active cochlear models for vowel recognition \cite{koizumi1996speech,ting2004speaker}, but they either predefine the gain factor in the active feedback, or use the automatic gain control proposed by Lyon \cite{lyon2011cascades} which introduces distortion artifacts \cite{ni2014modelling}. Most recently, the use of a convolutional neural network (CNN) structure to approximate the computations performed by the cascaded active cochlear models was proposed \cite{baby2020convolutional}, by training the neural network using data generated from a pre-existing active cochlear model. However, the network architecture is still somewhat arbitrary and more importantly lacks interpretability. In order to bridge the gap between computational models of the cochlea and realisable front-ends for machine based speech and audio processing tasks, we focus on developing realistic and jointly discrete cochlear models which can be implemented as discrete-time signal processing systems.

In this paper we propose a state space (SS) formulation of an active cochlear model that is jointly discretised in both time and space, which significantly reduces the computation cost and makes it a feasible front-end for speech processing systems. Further, the active mechanism of the cochlea is flexibly incorporated in the proposed approach, based on similar assumptions to those \cite{neely1986model,elliott2007state}. The cochlear responses to different input stimuli at different sound pressure levels (SPLs) are validated and compared with a semi-discretised model (spatially discrete). Further, the proposed model is analysed in terms of the key characteristics including the dynamic compression and system stability. Finally, the model response to speech signals is analysed to ascertain its effectiveness in capturing high-level speech representations for potential use in computational speech processing systems. 

The paper is organized as follows: Section~\ref{sec:2} briefly discusses the existing cochlear models with a focus on a semi-discretised cochlear model with active mechanism in SS form that forms the baseline system on which the proposal builds. Section~\ref{sec:3} introduces the proposed joint spatio-temporally discretised cochlear model. In Section~\ref{sec:4} we present experimental analyses of the cochlear model responses to different stimuli; and and Section~\ref{sec:5} discusses system stability and dynamic compression characteristics of the model. Finally, Section \ref{sec:6} explores the use of the proposed cochlear model for speech analyses.

\begin{figure}
\includegraphics[width=\reprintcolumnwidth]{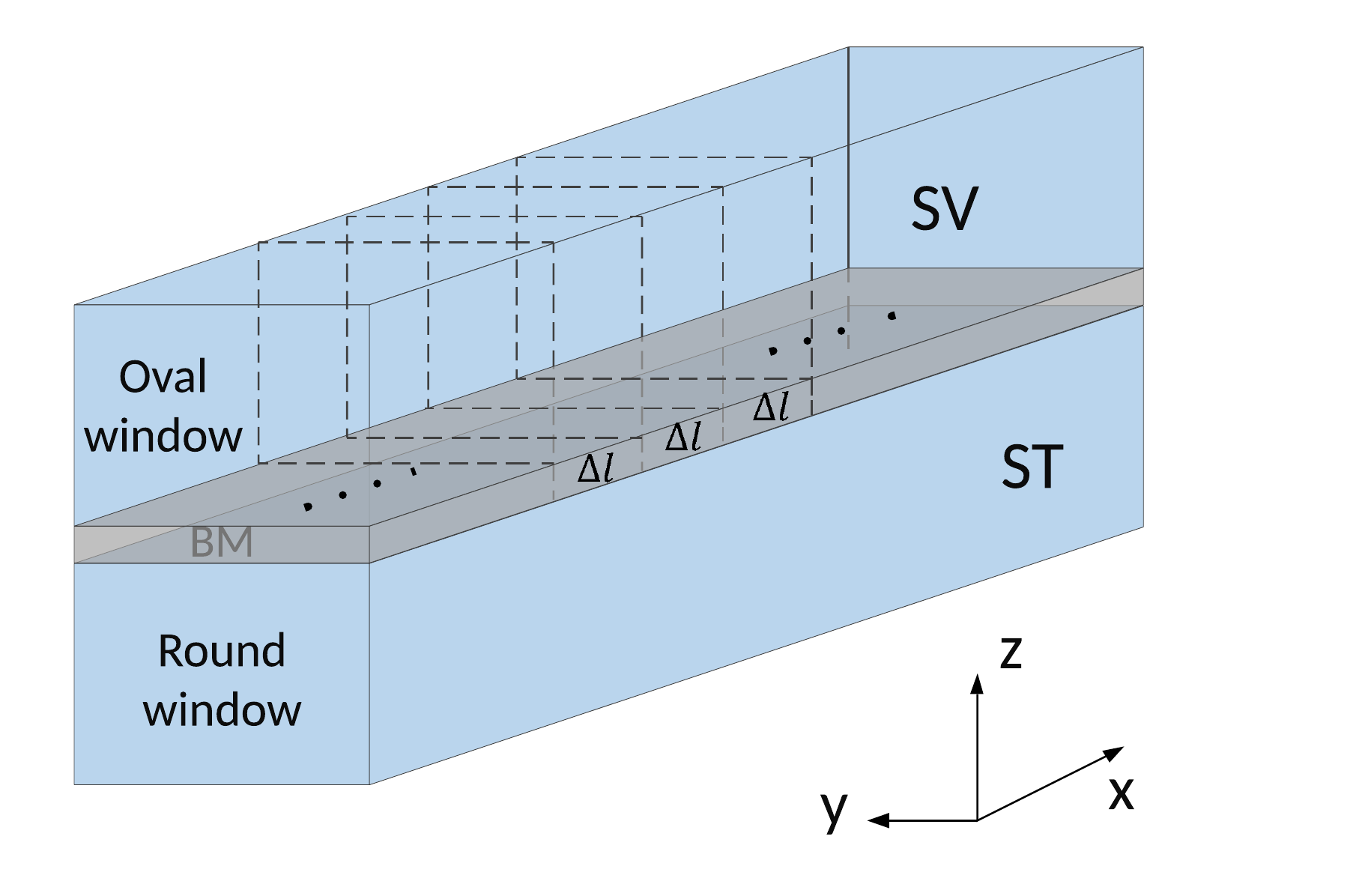}
\caption{\label{fig:real1}{A simplified 1-dimension box model for the cochlear. The upper chamber scala vestibuli (SV) and lower chamber of scala tympani (ST) are filled with fluid and separated by the Basilar Membrane (BM). The input stimulus  propagates through the oval window, and the pressure difference between upper and lower chambers generates BM vibrations. The finite element approach for the box model of the cochlear divides the BM into N elements with spatial difference $\Delta l$.}}
\raggedright
\end{figure}
 
\section{\label{sec:2} Background on cochlear models}
\subsection{Overview}
The cochlea is a spiral-shaped cavity in the bony labyrinth that is a part of the inner ear. Mathematical models of the cochlea can be broadly divided into three categories: (i) 1-dimensional models that only consider the transverse wave propagation in the longitudinal direction \cite{de1996mechanics,diependaal1989time}; (ii) 2-dimensional models that neglect the height of the cochlear chambers but consider both the transverse and radial wave propagation \cite{steele1979comparison, diependaal1989nonlinear,neely1981finite}; and (iii) 3-dimensional models that take into account all spatial constraints of the cavity \cite{steele1979comparison1,elliott2018elemental}. Among these, 1-dimensional models are the most common as they are computationally straightforward while still exhibiting very similar cochlear responses to the more complex 2D and 3D models in terms of the spectral characteristics of the cochlear responses \cite{de1996mechanics}, which is the primary characteristic of interest when designing front-ends for speech processing systems.

The 1-dimensional model in its simplest form is obtained as a box model \cite{de1996mechanics} with an upper fluid chamber referred to as the scala vestibuli (SV) and a lower fluid chamber referred to as the scala tympani (ST), which are separated by the Basilar Membrane (BM). The simplified structure is shown in Figure \ref{fig:real1}. This simple model is able to replicate the basic functions of the cochlea. Excited by the incoming sound wave at the oval window which lies at one end of the SV, the pressure difference between the upper and lower chamber generates a wave that travels along the entire duct of the box model. The characteristics of the cochlea also lead to tonotopy, i.e., different positions along the BM vibrate at a different characteristic frequency with the basal end corresponding to high frequencies and the apical at low frequencies.

This simple model however ignores a key element of the mammalian cochlea. Namely, that the SV and ST are separated by two membranes (not one), the aforementioned Basilar Membrane and the Tectorial Membrane (TM), that move relatively independently but are mechanically coupled via the surrounding fluid and the outer hair cells (OHCs). This coupled system involves a nonlinear active feedback mechanism (via the OHCs) that increases the frequency selectivity of tonotopic response of the BM to sound excitation, in addition to increasing the dynamic range of the input audio that the system can adequately respond to \cite{rhode1971observations,robles1976transient,johnstone1986basilar,ruggero1991furosemide,ruggero1992responses}. The nonlinearity manifests as (i) a higher gain at low input stimuli levels and a lower gain at high input stimuli levels, providing a form of automatic gain control \cite{lyon1988cochlear,lyon1990automatic}; (ii) a frequency sharpening mechanism that increases selectivity \cite{johnstone1986basilar,ruggero1991furosemide,neely1983active}; and (iii) vibrations in the BM and TM at positions corresponding to frequencies not present in the input stimuli \cite{wilson1980evidence}.  This nonlinear feedback mechanism underpins the ability of  mammals to perceive sounds over an extremely large dynamic range by allowing it to adapt to the input stimuli and there exists a body of research that has focused on developing nonlinear active mechanical models of the cochlea \cite{allen2001nonlinear,diependaal1989nonlinear,neely1986model}. These forms of active mechanisms are also absent in current front-ends for computational speech processing systems. 

An important class of 1D cochlear models involves a state space (SS) formulation that describes the dynamics of a nonlinear mechanical cochlear model as a set of coupled first-order differential equations \cite{elliott2007state}. This approach follows on from one of the most widely adopted 1D models proposed by Neely et al. in 1986 \cite{neely1986model}. The set of coupled first-order differential equations is obtained by spatially discretising the BM and TM into micro-elements, with each equation describing the mechanics of one of these segments.


\begin{figure*}[ht!]
\includegraphics[width=1\textwidth]{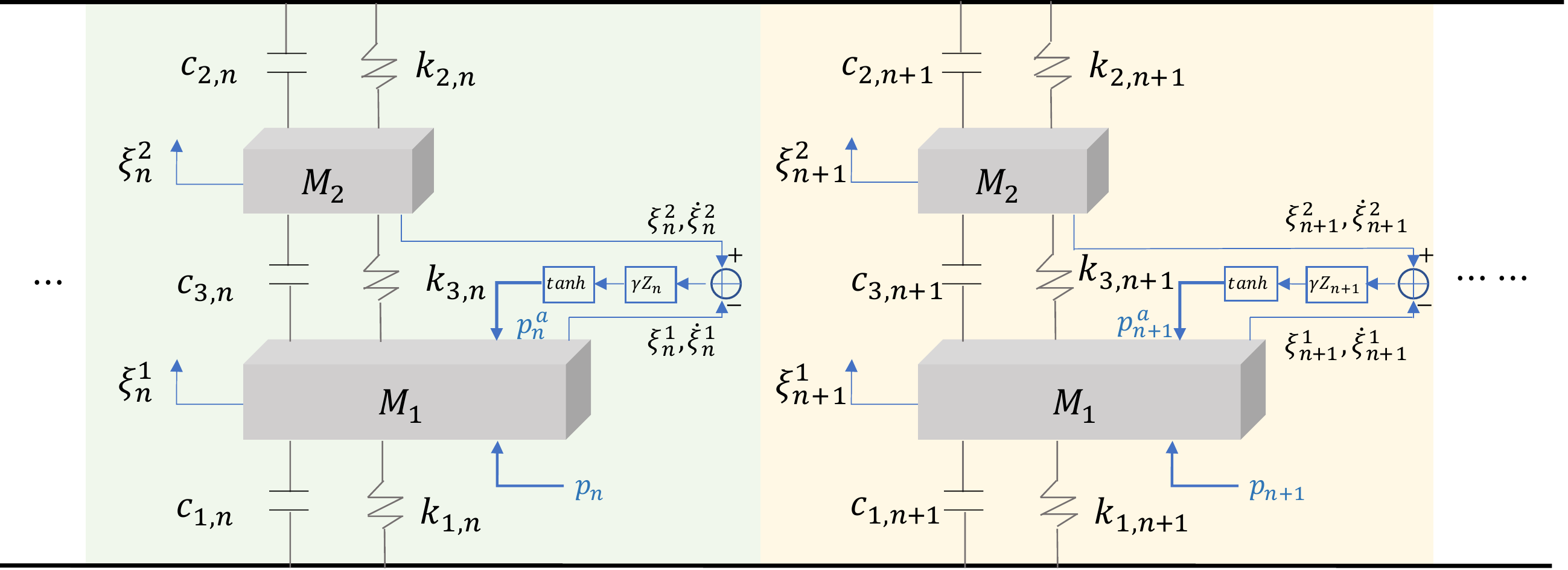}
\caption{The cascaded micro-mechanical model first proposed in \cite{neely1986model}. Each colored block represents one micro-element, which contains a section of the BM and TM represented by $M_1$ and $M_2$ respectively. The stiffness and damping coefficients are represented as $c_i$ and $k_i$ with $i$ of 1,2, and 3 indicating BM, TM and fluid coupling between BM and TM. $P^a$ is the active feedback simulating the OHC functions.}
\label{fig:real2}
\end{figure*}

\subsection{State space cochlear models}
A state space model of the cochlea comprises two linked components, one which describes the propagation of a pressure wave along the cochlea (referred to as macro-mechanics), and a second that describes the motion of sections of the cochlea in response to the pressure acting on it (referred to as micro-mechanics) as shown in Figure \ref{fig:real2}.


Based on the 1D box model, under the long wavelength assumption \cite{de1996mechanics}, the macro-mechanical model which describes wave propagation along the cochlea is given by the following equation:
\begin{equation}\label{eq:1}
\frac{\partial{p^2(l,t)}}{\partial{l^2}}=\frac{2\rho}{H}\ddot{\xi}(l,t)
\end{equation}
\noindent
where $p(l,t)$ represents the pressure difference between the upper and lower chambers of the cochlear at position $l$ and time $t$; $\ddot{\xi}(l,t)$ is the radially averaged transverse acceleration of the cochlear; $\rho$ is the fluid density of the chambers; and $H$ is the height of the cochlear chamber, which is assumed to be a constant.

Generally a finite difference approximation is used to spatially discretise the continuous wave propagation equation to obtain a semi-discretised description using $N$ small discrete elements, each of length $\Delta l$, along the spatial dimension $l$ to approximate the continuous variable. The resulting set differential equations \cite{neely1981finite} can be finally written in a matrix form as:
\begin{equation}\label{eq:2}
    \mathbf{Fp}(t) - {\mathbf{\ddot\xi}}(t) = \mathbf{q}(t)
\end{equation}
\noindent

\noindent
where $\mathbf{p}(t)$ is the $N$ dimensional vector of pressure differences across all elements at time $t$. $\ddot{\xi}(t)$ represents the acceleration of these elements, $\mathbf{q}(t)$ is an $N$ dimensional vector of external excitation terms corresponding to each element, and $\boldsymbol{F}$ is the finite-difference matrix given as:
\begin{equation}\label{eq:3}
\mathbf{F} = \frac{H}{2 \rho\Delta l^2}\begin{bmatrix} -\frac{\Delta l}{H}& \frac{\Delta l}{H} & 0 & 0 &  \dots & 0 \\ 1 & -2 & 1 & 0 & \dots & 0 \\ \vdots & \vdots & \vdots & \ddots & \vdots & \vdots \\0  & \dots & 1 & -2 & 1 & 0 \\ 0  & \dots & 0 & 0 & 0  & -\frac{2 \rho\Delta l^2}{H} \end{bmatrix}
\end{equation}

Since the only input to the cochlea is via the oval window only the first element is non-zero and $\mathbf{q}(t)$ is given as:
\begin{equation}\label{eq:4}
   \mathbf{q}(t) = [\ddot{\xi}_s,0,0,\cdots, 0]^T
\end{equation}

\noindent
where $\ddot{\mathbf{\xi}}_s(t)$ is the acceleration due to the loading by the internal pressure response at the basal end \cite{elliott2007state}. Mathematical derivations of these expressions can be found in \hyperref[app:A]{Appendix A}.

A widely adopted micro-mechanical model describing the mechanical responses within each discretised spatial segment is the 2-degree of freedom model with active feedback \cite{neely1986model}. The components of this micro-mechanical model of each segment and the cascaded coupling with the next segment is outlined in Figure \ref{fig:real2}. Each colored block represents one spatial segment. Here $M_1$ and $M_2$ represent the mass of the BM and TM segments respectively. The stiffness and damping coefficients associated with these masses are represented as $c_i$ and $k_i$ with $i = 1, 2,$ and $3$ denoting the BM, TM and fluid coupling between BM and TM. It should be noted that parameters of $c_i$ and $k_i$ vary with $l$ but are not denoted explicitly to simplify notation in the following discussions. An additional pressure term in each spatial segment $p_n^a$ is introduced to denote the active feedback via the outer hair cell (OHC) mechanical stimuli to the $n^{th}$ BM element. Within each segment, the micro-mechanics described by the two force equations corresponding to the BM and TM are given by:

\begin{equation}\label{eq:5}
    p_n - p_n^a = \overline{\alpha}_n^{2} \ddot{\xi_n^1} + \overline{\alpha}_n^{1} \dot{\xi_n^1} + \overline{\alpha}_n^{0} \xi_n^1 + \overline{\beta}_n^{1} \dot{\xi_n^2} + \overline{\beta}_n^{0} \xi_n^2  
\end{equation}

\begin{equation}\label{eq:6}
    0 =  \overline{\delta}_n^{1} \dot{\xi_n^1} + \overline{\delta}_n^{0} \xi_n^1 + \overline{\varepsilon}_n^{2}\ddot{\xi_n^2}+\overline{\varepsilon}_{n}^1 \dot{\xi_n^2} + \overline{\varepsilon}_n^{0} \xi_n^2
\end{equation}
where $\xi_n^1$ and $\xi_n^2$ represent the displacement of the BM and TM in the $n_{th}$ segment, $\dot{\xi_n^1}$ and $\dot{\xi_n^2}$ represent the corresponding velocities, and $\ddot{\xi_n^1}$ and $\ddot{\xi_n^2}$ the accelerations. The active feedback based on Neely's model is defined in terms of BM and TM displacements and velocities, and introduced in equation (\ref{eq:5}) by altering the model parameters. Details are discussed in section \ref{sec:3}. The derivations of equations (\ref{eq:5}) and (\ref{eq:6}) and parameter values of $\overline{\alpha}$, $\overline{\beta}$, $\overline{\delta}$ and $\overline{\varepsilon}$ are provided in \hyperref[app:B]{Appendix B}.

The authors\cite{elliott2007state} formulate a state-space model that integrates the macro- and micro-mechanical models, by introducing a state vector describing the state of all $N$ segments, $\mathbf{x}(t) = [\mathbf{x}_1^T(t),\mathbf{x}_2^T(t),\cdots, \mathbf{x}_N^T(t)]^T$, with each element $\mathbf{x}_n(t)$ comprising four state variables, namely, the velocity and displacement of the BM and TM within each segment. i.e.,
\begin{equation}\label{eq:7}
 \mathbf{x}_n(t) =[\dot{\xi_n^1}(t),\xi_n^1(t),\dot{\xi_n^2}(t),\xi_n^2(t)]^T
\end{equation}
\noindent
which leads to the following state-space equation:
\begin{equation}\label{eq:8}
    \Dot{\mathbf{x}}(t)=\mathbf{A}\mathbf{x}(t)+\mathbf{B}\mathbf{u}(t)
\end{equation}
\noindent
where the system matrices $\mathbf{A}$ and $\mathbf{B}$ represent the mechanical properties within each segment as per the micro-mechanical model and the pressure generated by the travelling wave as per the macro-mechanical model\cite{elliott2007state}. A brief mathematical derivation of the state-space formulation is provided in \hyperref[app:C]{Appendix C}.

Numerical ODE solvers can be used to solve this semi-discretised (spatially discrete but temporally continuous) state space equation (\ref{eq:8}) for BM and TM displacements (and velocities). However, most ODE solvers employ adaptive time steps when obtaining these solutions which is in contrast to typical discrete signal processing (DSP) systems that operate at regular time steps. Therefore we aim to develop a jointly discrete model (both spatially and temporally discrete) that allows for the BM and TM displacements to be obtained with fixed time step computations.

\section{Proposed joint discretized model\label{sec:3}}
The proposed joint discretised model is obtained by employing a first order finite difference method in both temporal and spatial domains, which leads to a relatively straightforward jointly discrete formulation. Following this, implementation of the nonlinear feedback in the cochlea and the stability of the proposed model are discussed.


\subsection{Macromechanics}
Discretising the wave propagation equation (\ref{eq:1}) using Euler's method on both the time and position variables leads to: 
\begin{equation}\label{eq:9}
\begin{split}
      & \frac{p_{j,n+1}-2p_{j,n}+p_{j,n-1}}{\Delta l^2}=  \\
      &\frac{2\rho}{H}\frac{\xi_{j+1,n}^1-2\xi_{j,n}^1+\xi_{j-1,n}^1}{\Delta {t}^2} + \frac{2\rho}{H} q_{j,n}
\end{split}
\end{equation}
\noindent
where $\Delta {l}$ denotes the length of each of the $N$ spatial segments, $\Delta {t}$ denotes the duration of each discretised time step, $p_{j,n}$ and $\xi_{j,n}^1$ represent the pressure and BM displacement at the $n_{th}$ element at the $j^{th}$ time step (i.e., at $t = j\Delta {t}$), which evolve in response to excitation of the $q_{j,n}$. This excitation is non-zero only at $n=1$ (the first discrete element of the cochlear model), where it corresponds to the incident pressure changes in the ear canal. i.e., $q_{j,n}=0$ when $n \neq 1$.


This discretisation leads to a set of $N$ linear equations, one for each spatial element describing the time evolution of the BM displacement in that section. Taking into account the boundary conditions, and aiming to formulate a state space representation similar to equation (\ref{eq:8}), this set of linear equations can be represented as a matrix equation:
\begin{equation}\label{eq:10}
   \Delta{t}^2\mathbf{F}\mathbf{P}_j=\mathbf{S}_2(\mathbf{X}_{j+1}-2\mathbf{X}_j+\mathbf{X}_{j-1}) + \Delta{t}^2\mathbf{Q}_{j}
\end{equation}
\noindent
where $\mathbf{F}$ is the finite-difference matrix as shown in equation (\ref{eq:3}), $\mathbf{P}_j$, the pressure vector comprising $N$ elements is:
\begin{equation}\label{eq:11}
\mathbf{P}_j = [p_{j,1},p_{j,2},\cdots,p_{j,N}]^T
\end{equation}
\noindent
and the state vector $\mathbf{X}_j$ including both BM and TM displacements over $N$ elements is:
\begin{equation}\label{eq:12}
\mathbf{X}_j = \begin{bmatrix} \xi_{j,1}^1, & \xi_{j,1}^2, \cdots & \xi_{j,n}^1, & \xi_{j,n}^2,  & \cdots & \xi_{j,N}^1, & \xi_{j,N}^2 \end{bmatrix}^T
\end{equation}
\noindent
with $\xi_{j,n}^1$ and $\xi_{j,n}^2$ representing the $n^{th}$ BM and TM displacements at time $j$, and $\mathbf{S}_2$ denoting the $N$ x $2N$ matrix that only selects the BM elements:
\begin{equation}\label{eq:13}
\mathbf{S}_2 = \begin{bmatrix} 1& 0 & 0 & 0 &  \dots & 0 \\ 0 & 0 & 1 & 0 & \dots & 0 \\ \vdots & \vdots & \vdots & \ddots & \vdots & \vdots \\ 0  & \dots & 0 & 0 & 1  & 0 \end{bmatrix}
\end{equation}
\noindent
The excitation (input stimuli) is given by the $N$ dimensional vector $\mathbf{Q}_j$:
\begin{equation}\label{eq:14}
\mathbf{Q}_j = \begin{bmatrix} q_{j,1}, & 0, & 0, & 0,  &  \cdots & 0, & 0 \end{bmatrix}^T
\end{equation}
\noindent

\subsection{Micromechanics}\label{sec:3.2}
As shown in Figure \ref{fig:real2}, which depicts the micro-mechanical model, the additional pressure term, $p_a(l,t)$, operating on each segment of the BM is used to simulate the impact of OHCs which introduces active (nonlinear) feedback. As previously discussed, the action of healthy OHCs are able to amplify the BM response to input stimuli at small SPLs and in turn lead to significant dynamic range compression. In this section, we first describe a linear feedback mechanism and then introduce the necessary nonlinearity into this model in section \ref{sec:3.5}.

The linear active feedback in the widely adopted Neely model is given by:
\begin{equation}\label{eq:15}
    p_a(l,t) = - \gamma(c_4\dot{\xi^f}(l,t)+k_4\xi^f(l,t))
\end{equation}
where

\begin{equation}\label{eq:16}
    \xi^f(l,t)=g\xi^1(l,t)-\xi^2(l,t)
\end{equation}
represents the relative displacement between the TM and Reticular Lamina (RL) where the displacement of the RL is known to be proportional to the displacement of the BM with a proportionality constant $g$, $c_4$ and $k_4$ represent the damper and stiffness coefficients that control the feedback terms that are position-dependent parameters, and $\gamma$ is the feedback gain. 

The proposed discretisation of the active feedback term $p_a(l,t)$ leads to:
\begin{equation}\label{eq:17}
    p^a_{j,n} = -\gamma (c_4 \frac{\xi^f_{j+1,n} -\xi^f_{j,n} }{\Delta_{t}} +k_4\xi^f_{j,n})
\end{equation}

Discretising the force equations of (\ref{eq:5}) and (\ref{eq:6}) and integrating the discretised feedback term from equation (\ref{eq:17}) leads to the final jointly discretised micro-mechanical model:
\begin{equation}\label{eq:18}
\begin{split}
      p_{j,n} = & \alpha_{n}^1 \xi_{j+1,n}^{1} + \alpha_{n}^0 \xi_{j,n}^{1} + \alpha_{n}^{-1}\xi_{j-1,n}^{1}  \\
  & +\beta_{n}^1 \xi_{j+1,n}^{2} + \beta_{n}^0 \xi_{j,n}^{2}
\end{split}
\end{equation}

\begin{equation}\label{eq:19}
\begin{split}
  0 = & \varepsilon_n^1 \xi_{j+1,n}^{1} + \varepsilon_n^0 \xi_{j,n}^{1} +  \\
  & +\delta_n^1 \xi_{j+1,n}^{2} + \delta_n^0 \xi_{j,n}^{2} + \delta_n^{-1} \xi_{j-1,n}^{2}
  \end{split}
\end{equation}


\noindent
where $\alpha^1_n$, $\alpha^0_n$, $\alpha^{-1}_n$, $\beta^1_n$, $\beta_n^0$, $\varepsilon_n^1$, $\varepsilon_n^0$, $\delta_n^1$, $\delta_n^0$ and $\delta_n^{-1}$ are model parameters corresponding to the $n^{th}$ segment. Expressions for these terms are provided in \hyperref[app:D]{Appendix D}.

The two force equations for the $n_{th}$ element of BM and TM can therefore be rearranged as:
\begin{equation}\label{eq:20}
\begin{split}
\begin{bmatrix}
p_{j,n} \\ 0 
\end{bmatrix} & =
\begin{bmatrix} \alpha_n^1 & \beta_n^1 \\  \varepsilon_n^1 & \delta_n^1 \end{bmatrix} 
\begin{bmatrix} \xi_{j+1,n}^1 \\ \xi_{j+1,n}^2 \end{bmatrix}+
\begin{bmatrix} \alpha_n^0 & \beta_n^0 \\  \varepsilon_n^0 & \delta_n^0 \end{bmatrix}
\begin{bmatrix} \xi_{j,n}^1 \\ \xi_{j,n}^2 \end{bmatrix}  \\ 
& + \begin{bmatrix} \alpha_n^{-1} & 0 \\ 0 & \delta_n^{-1} \end{bmatrix}
\begin{bmatrix} \xi_{j-1,n}^1 \\ \xi_{j-1,n}^2 \end{bmatrix}
\end{split}
\end{equation}
and the jointly discretised micro-model over all $N$ BM elements can then be represented as:
\begin{equation}\label{eq:21}
\mathbf{\Omega}_j = \mathbf{A}_1 \mathbf{X}_{j+1} + \mathbf{A}_0 \mathbf{X}_j + \mathbf{A}_{-1} \mathbf{X}_{j-1}
\end{equation}
where $\mathbf{\Omega}_j$ is given by:
\begin{equation}\label{eq:22}
    \mathbf{\Omega}_j = [p_{j,1}, 0, p_{j,2}, 0 , \cdots , p_{j,N}, 0]^T
\end{equation}
and $\mathbf{X}_j$ is a $2N$ x $1$ vector that represents the state of the BM and TM at time step $j$ and is given as per equation (\ref{eq:12}). The matrices $\textbf{A}_1$, $\textbf{A}_0$ and $\textbf{A}_{-1}$ are block-diagonal matrices that hold the micro-mechanical model parameters:

\begin{equation}
     \mathbf{A}_1 = \begin{bmatrix}
   \alpha_1^1 & \beta_1^1 & 0 & 0 & \cdots & 0 & 0 \\ \varepsilon_1^1 & \delta_1^1 & 0 & 0 & \cdots & 0 & 0 \\
   0 & 0 & \alpha_2^1 & \beta_2^1 & \cdots & 0 & 0  \\ 
   0 & 0 &  \varepsilon_2^1 & \delta_2^1 & \cdots & 0 & 0  \\ 
   \vdots & \vdots & \cdots & \cdots & \ddots & \vdots & \vdots \\
   0 & 0 & 0 & 0 & \cdots & \alpha_N^1 & \beta_N^1 \\0 & 0 & 0 & 0 & \cdots & \varepsilon_N^1 & \delta_N^1 
   \end{bmatrix}
\end{equation}

\begin{equation}
     \mathbf{A}_0 = \begin{bmatrix}
   \alpha_1^0 & \beta_1^0 & 0 & 0 & \cdots & 0 & 0 \\ \varepsilon_1^0 & \delta_1^0 & 0 & 0 & \cdots & 0 & 0 \\
   0 & 0 & \alpha_2^0 & \beta_2^0 & \cdots & 0 & 0 \\ 
   0 & 0 &  \varepsilon_2^0 & \delta_2^0 & \cdots & 0 & 0 \\ 
   \vdots & \vdots & \cdots & \cdots & \ddots & \vdots & \vdots \\
   0 & 0 & 0 & 0 & \cdots & \alpha_N^0 & \beta_N^0 \\0 & 0 & 0 & 0 & \cdots & \varepsilon_N^0 & \delta_N^0
   \end{bmatrix}
\end{equation}

\begin{equation}
     \mathbf{A}_{-1} = \begin{bmatrix}
   \alpha_1^{-1} & 0 & 0 & 0 & \cdots & 0 & 0 \\ 0 & \delta_1^{-1} & 0 & 0 & \cdots & 0 & 0 \\
   0 & 0 & \alpha_2^{-1} & 0 & \cdots & 0 & 0  \\ 
   0 & 0 &  0 & \delta_2^{-1} & \cdots & 0 & 0  \\ 
   \vdots & \vdots & \cdots & \cdots & \ddots & \vdots & \vdots \\
   0 & 0 & 0 & 0 & \cdots & \alpha_N^{-1} & 0 \\0 & 0 & 0 & 0 & \cdots & 0 & \delta_N^{-1}
   \end{bmatrix}
\end{equation}

\subsection{Numerical solutions for joint discretized models}
To infer the BM and TM displacements we solve for $\mathbf{X}_{j}$ at each time step $j$ by combining the macro-mechanical and micro-mechanical models given by equations (\ref{eq:10}) and (\ref{eq:21}) respectively. Noting that we can rewrite (\ref{eq:10}) as:
\begin{equation}\label{eq:26}
    \mathbf{P}_j = \frac{1}{\Delta {t}^2} \mathbf{F}^{-1} \mathbf{S}_2 (\mathbf{X}_{j+1}-2\mathbf{X}_j+\mathbf{X}_{j-1}) + \mathbf{F}^{-1} \mathbf{Q}_j
\end{equation}
and that,
\begin{equation}\label{eq:27}
   \mathbf{\Omega}_j = \mathbf{S}_2^T \mathbf{P}_j   
\end{equation}
\noindent
where $\mathbf{S}_2^T$ is the transpose of $\mathbf{S}_2$ given as per (\ref{eq:13}), the interaction between the macro- and micro-mechanical models can then be represented as:
\begin{equation}\label{eq:28}
\begin{split}
\mathbf{S}_2^T \mathbf{P}_j   &= \mathbf{A}_1 \mathbf{X}_{j+1} + \mathbf{A}_0 \mathbf{X}_j + \mathbf{A}_{-1} \mathbf{X}_{j-1} \\ & =\mathbf{\Gamma} (\mathbf{X}_{j+1}-2\mathbf{X}_j+\mathbf{X}_{j-1}) + \mathbf{S}_2^T \mathbf{F}^{-1} \mathbf{Q}_j 
\end{split}
\end{equation}
where $\mathbf{\Gamma}$ is:
\begin{equation}\label{eq:29}
    \mathbf{\Gamma} = \frac{1}{\Delta_{t}^2} \mathbf{S}_2^T  \mathbf{F}^{-1} \mathbf{S}_2
\end{equation} 
Consequently, we can solve for $\mathbf{X}_j$ as: 
\begin{equation}\label{eq:30}
    \mathbf{X}_{j+1}  = \mathbf{H}\mathbf{X}_{j} 
    + \mathbf{K} \mathbf{X}_{j-1} 
 + \mathbf{M}\mathbf{U}_j
\end{equation}
\noindent
where,
\begin{align}
    &\mathbf{H}= (\mathbf{A}_1- \mathbf{\Gamma})^{-1} (-2\mathbf{\Gamma} - \mathbf{A}_0) \\
&\mathbf{K}= (\mathbf{A}_1- \mathbf{\Gamma})^{-1}(\mathbf{\Gamma} - \mathbf{A}_{-1}) \\
 &\mathbf{M}= (\mathbf{A}_1- \mathbf{\Gamma})^{-1}\mathbf{S}_2^T \\
&\mathbf{U}_j =  \mathbf{F}^{-1} \mathbf{Q}_j
\end{align}

Note that $\mathbf{U}_j$ captures the input signal/excitation and is consistent with \cite{elliott2007state}. It is clear that the BM and TM displacements are computed iteratively depending on the state at the previous two time steps and the current state of the input signal.

In a linear active cochlear model with a constant $\gamma$ governing the feedback term, the matrices $\mathbf{H}$, $\mathbf{K}$ and $\mathbf{M}$ are all fixed, and (\ref{eq:30}) represents a linear discrete-time state space representation of the cochlear model.

\subsection{System stability}\label{sec:3.4}
To analyse the stability of this jointly discretised model, we recast the model in terms of an alternative state vector, $\mathbf{Y}_{j}= [\mathbf{X}_{j},\mathbf{X}_{j-1}]^T$, and rewrite equation (\ref{eq:30}) as:
\begin{equation}\label{eq:35}
 \mathbf{Y}_{j+1} =  \mathbf{E} \mathbf{Y}_{j}  + \begin{bmatrix} \mathbf{M}\mathbf{U}_j \\ \mathbf{0} \end{bmatrix}
\end{equation}
\noindent
where $\mathbf{E}$ is given as:
\begin{equation}\label{eq:36}
 \mathbf{E}= \begin{bmatrix}
 \mathbf{H} & \mathbf{K} \\ \mathbf{I}_{2N} & \mathbf{0}
 \end{bmatrix} 
\end{equation}

Iterating backwards to the first time step, it can be seen that $\mathbf{Y}_{j+1}$ can also be written as:
\begin{equation}\label{eq:37}
    \mathbf{Y}_{j+1} =  \mathbf{E}^j \mathbf{Y}_1 + \begin{bmatrix} \sum_{k=0}^{j-1}\mathbf{E}^k\mathbf{M}\mathbf{U}_{j-k} \\ \mathbf{0} \end{bmatrix}
\end{equation}
where $\mathbf{E}^k = \underbrace{\mathbf{E} \mathbf{E} \cdots \mathbf{E}}_{k \; times}$.

Consequently, it is clear that the stability of the system is determined by the eigenvalues of $\mathbf{E}$. Namely, the magnitude of all its eigenvalues must be less than $1$. 

The only two parameters involved in the spatio-temporal discretisation are the length of the spatial segments $(\Delta {l})$ and the duration of each time step $(\Delta {t})$. In the proposed model $\Delta {l}$ is chosen to match \cite{elliott2007state}, and consequently stability is determined by $\Delta {t}$. In section \ref{sec:stability_analyses} we present a brief analysis of the choice of $\Delta {t}$ that leads to stable models.



\subsection{Nonlinear feedback}\label{sec:3.5}
As previously mentioned, the feedback via OHCs in the cochlea is not linear and an established approach to introduce this nonlinearity in cochlear models is by including a nonlinear tanh function in the otherwise linear feedback path \cite{neely1986model}. As tanh is fairly linear for small inputs, it still provides a similar level of amplification to inputs with small SPLs as in the linear active model. However, for large inputs, the tanh function saturates, which in turn saturates the gain. This property leads to the desirable dynamic range compression. 

The nonlinear feedback term $p_{j,n}^a$ can thus be represented as:
\begin{equation}\label{eq:38}
p_{j,n}^{a'} = \mathrm{tanh}( \tau p_{j,n}^{a})  
\end{equation}
\noindent
where $\tau$ is a scalar that can be used to control the saturation characteristics of the nonlinearity. 

\begin{figure}[b!]
    \centering
    \includegraphics[width=0.5\textwidth]{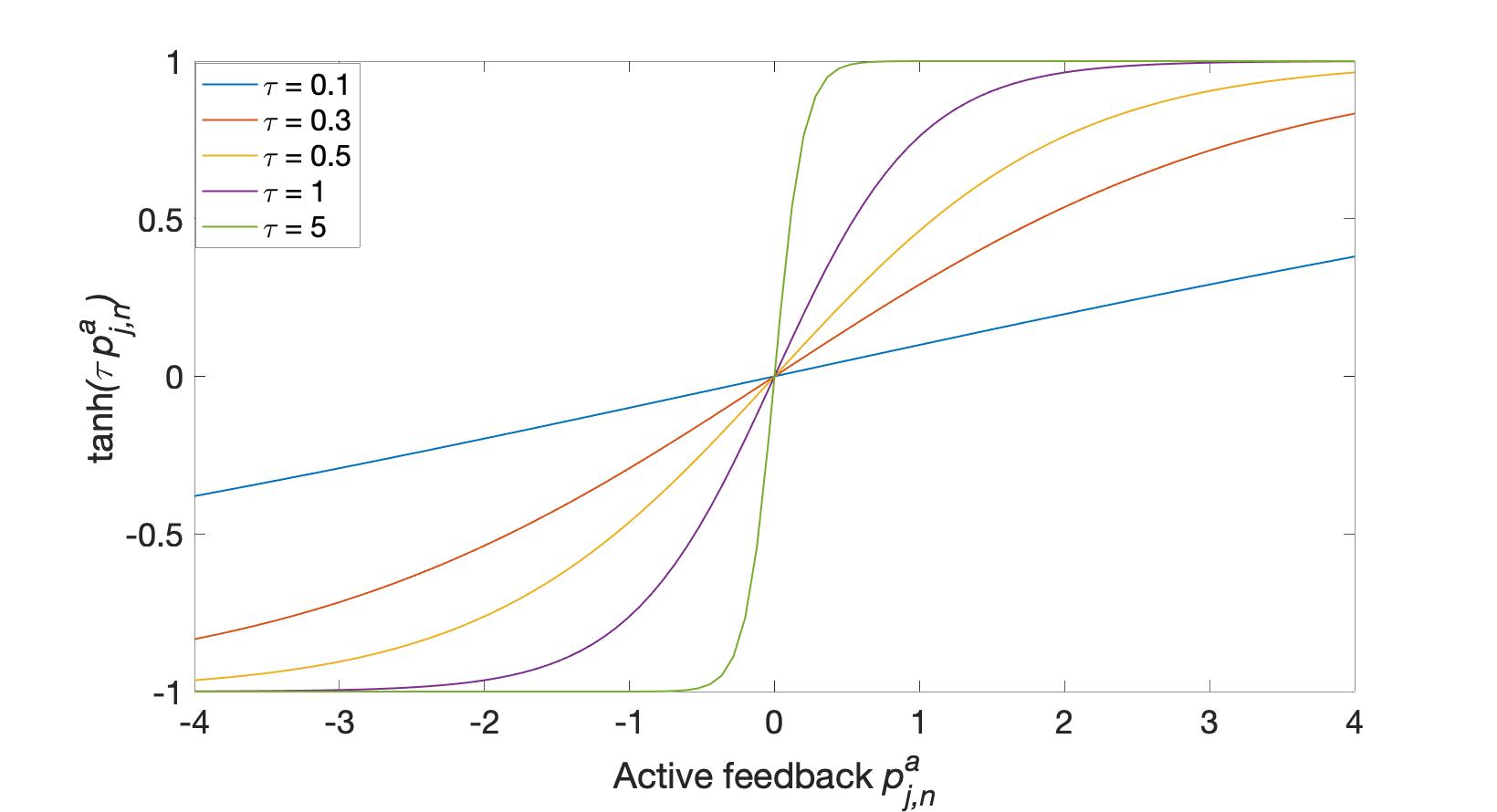}
    \caption{The impact of different $\tau$ values in the active feedback. A larger $\tau$ value leads to a saturation point at smaller additional pressure $p_{j,n}^{a}$, which indicates a more significant compression.}
    \label{fig:beta}
\end{figure}

The impact of different $\tau$ values is shown in Figure \ref{fig:beta}. With $\tau$ value increasing from 0.1 (light blue line) to 5 (cyan line), the saturation point with respect to active feedback $p_{j,n}^{a}$ becomes smaller, indicating that saturation starts to function at a smaller input SPL, achieving a more significant compression capability. To achieve a large dynamic range of compression, a larger $\tau$ is preferred. However, $\tau$ larger than 1 results in an unstable system. Therefore,  we adopt $\tau=1$ for all the following experiments, aiming for a high compression capability while processing speech data to decrease the variations in speech representations as well as improving the robustness to the noise in the input.

To simplify the implementation, a time varying scaling factor $\omega_{j,n}$ is proposed to replace the $tanh$ impact as:
\begin{equation}\label{eq:39}
\omega_{j,n} =\frac{\mathrm{tanh}(p_{j,n}^{a}) }{p_{j,n}^{a}}
\end{equation}
It should be noted that $\omega_{j,n}$ is a time-varying factor which requires computation at each time step. 

Apart from the tanh function, we also explored the use of the first order Boltzmann function, which has more degrees of freedom, as the nonlinearity \cite{ku2008modelling}. However, preliminary results suggested that the improvement in the resulting dynamic range compression was very small and not likely to be worth the increase in complexity. We did not pursue it further.


\begin{table}[t]
    \caption{Parameter values used to implement the jointly discretised nonlinear cochlear model \cite{ku2008statistics}}
    \centering
    \begin{tabular}{ c|c } 
    \hline
    Quantity & Formula (SI)\\
    \hline
     $ m_1$ & $1.35 \times 10^{-2} \mathrm{kg  \:m^{-2}}$ \\
     $m_2$ & $2.3\times10^{-3} \mathrm{kg\: m^{-2}}$ \\
     $k_1$ & $4.95 \times 10^9e^{-320(x+0.00375)} \mathrm{N\: m^{-3}}$ \\
     $k_2$ & $3.15 \times 10^7e^{-352(x+0.00375)}\mathrm{N\: m^{-3}}$ \\
    $k_3$ & $4.5 \times 10^7e^{-320(x+0.00375)}\mathrm{N\: m^{-3}}$\\
     $k_4$ & $2.82\times10^9e^{-320(x+0.00375)}\mathrm{N\: m^{-3}}$\\
     $c_1$ & $1+19700e^{-179(x+0.00375)}\mathrm{N\: s\: m^{-3}}$ \\
     $c_2$ & $113e^{-176(x+0.00375)} \mathrm{N\: s\: m^{-3}}$\\
     $c_3$ & $22.5e^{-64(x+0.00375)}\mathrm{N\: s\: m^{-3}}$\\
     $c_4$ & $9650e^{-164(x+0.00375)}\mathrm{N\: s\: m^{-3}}$\\
     $m_{ME}$ & $2.96\times10^{-2} \mathrm{kg\: m^{-2}}$\\
     $k_{ME}$ & $2.63\times10^8 \mathrm{N\: m^{-3}}$\\
     $c_{ME}$ & $2.8\times10^4 \mathrm{N\: s\: m^{-3}}$\\
     $L$ & 0.035m \\
     $H$ & 0.0001m \\
     $N$ & 500 \\
     $\gamma$ & 1 \\
     \hline
    \end{tabular}
    \label{tab:real1}
\end{table}

\begin{figure}[t!]
    \centering
    \includegraphics[width=0.5\textwidth]{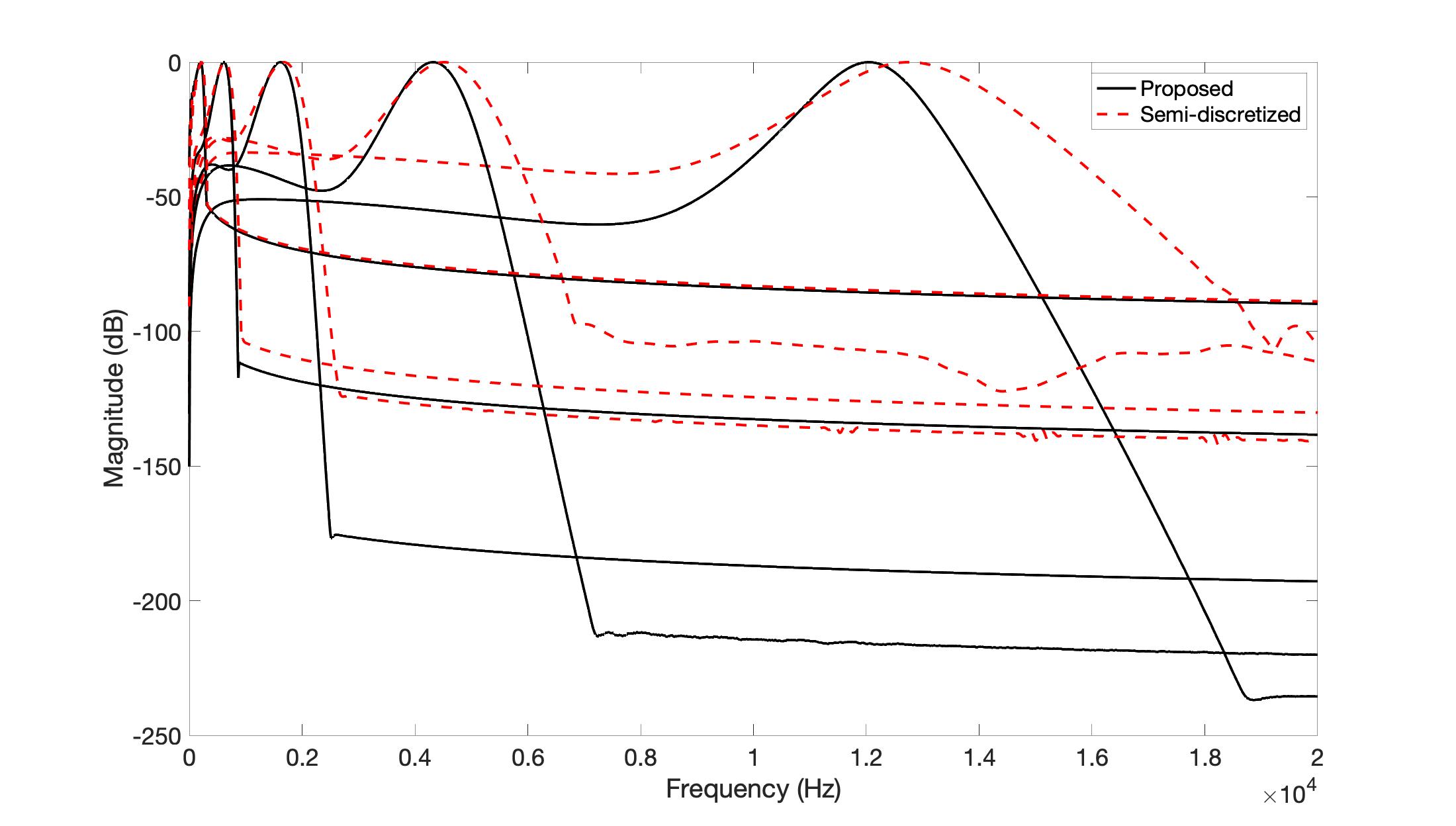}
    \caption{Magnitude responses of 'cochlear filters' corresponding to 5 randomly chosen BM elements estimated based on the response to a click (impulse) stimulus at 0dB (SPL). The magnitude responses are normalised such that the maximum is 0dB for each BM element. The solid black line shows the response from the proposed model while the red dashed line is that of the same element from the semi-discretised model. A similar pattern is observed in both systems. The discrepancy between two systems in high frequency is due to the temporal discretisation, and they are aligned well with a smaller time step (a larger sampling frequency).}
    \label{fig:2}
\end{figure}

\section{\label{sec:4} Validation of Model Responses}
The jointly discretised model described in this paper was validated by analysing the BM responses given by the model in response to given different types of input stimulus, including an impulse signal, single tone sinusoids, and sinusoids with time-varying frequencies. The parameter values (in SI units) utilised when implementing are listed in Table 1. These are identical to those used in \cite{ku2008statistics}.

\begin{figure*}[ht!]
\parskip=3pt
\baselineskip=6pt
\noindent
\figline{\fig{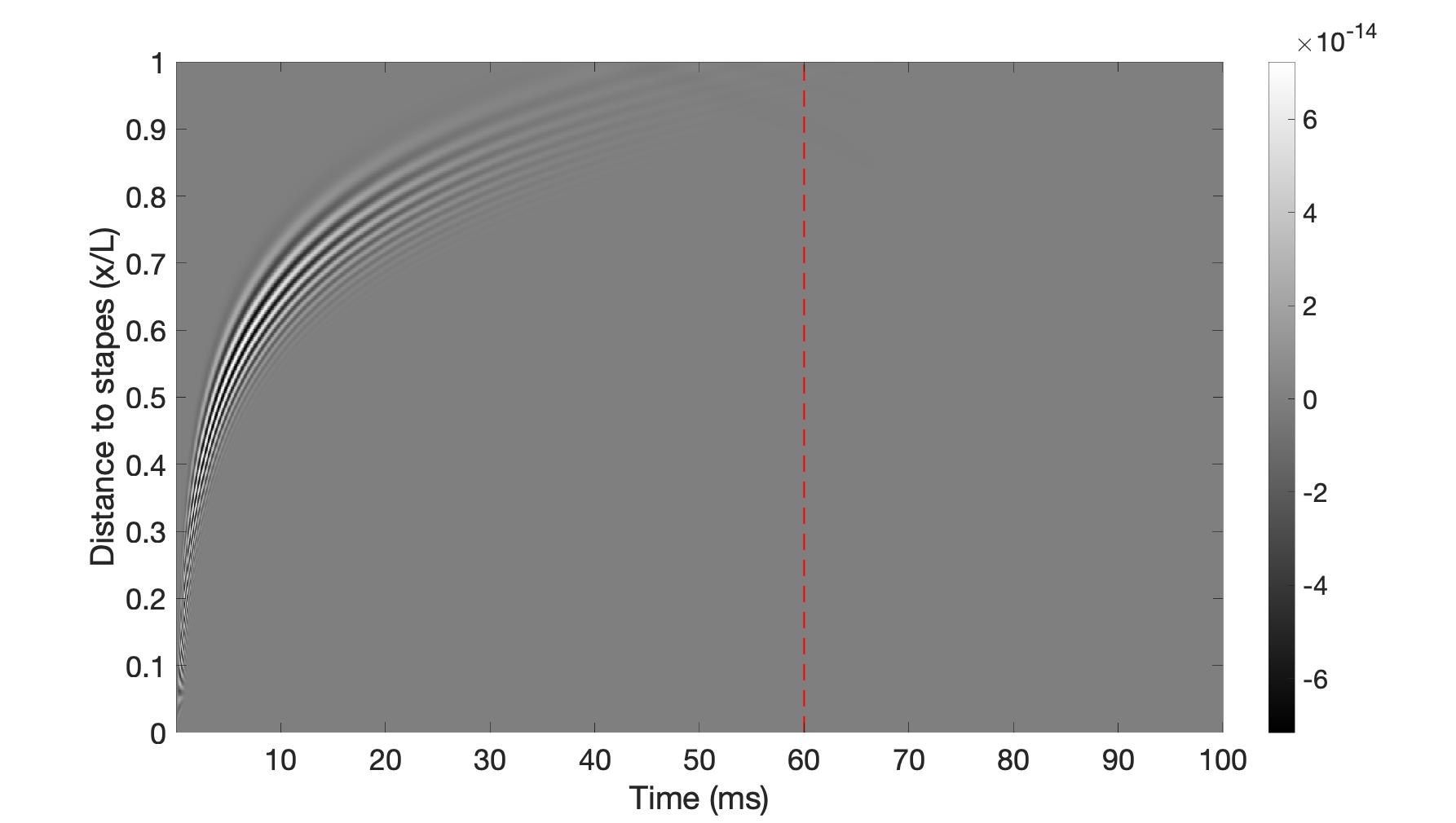}{9cm}{(a)}
\fig{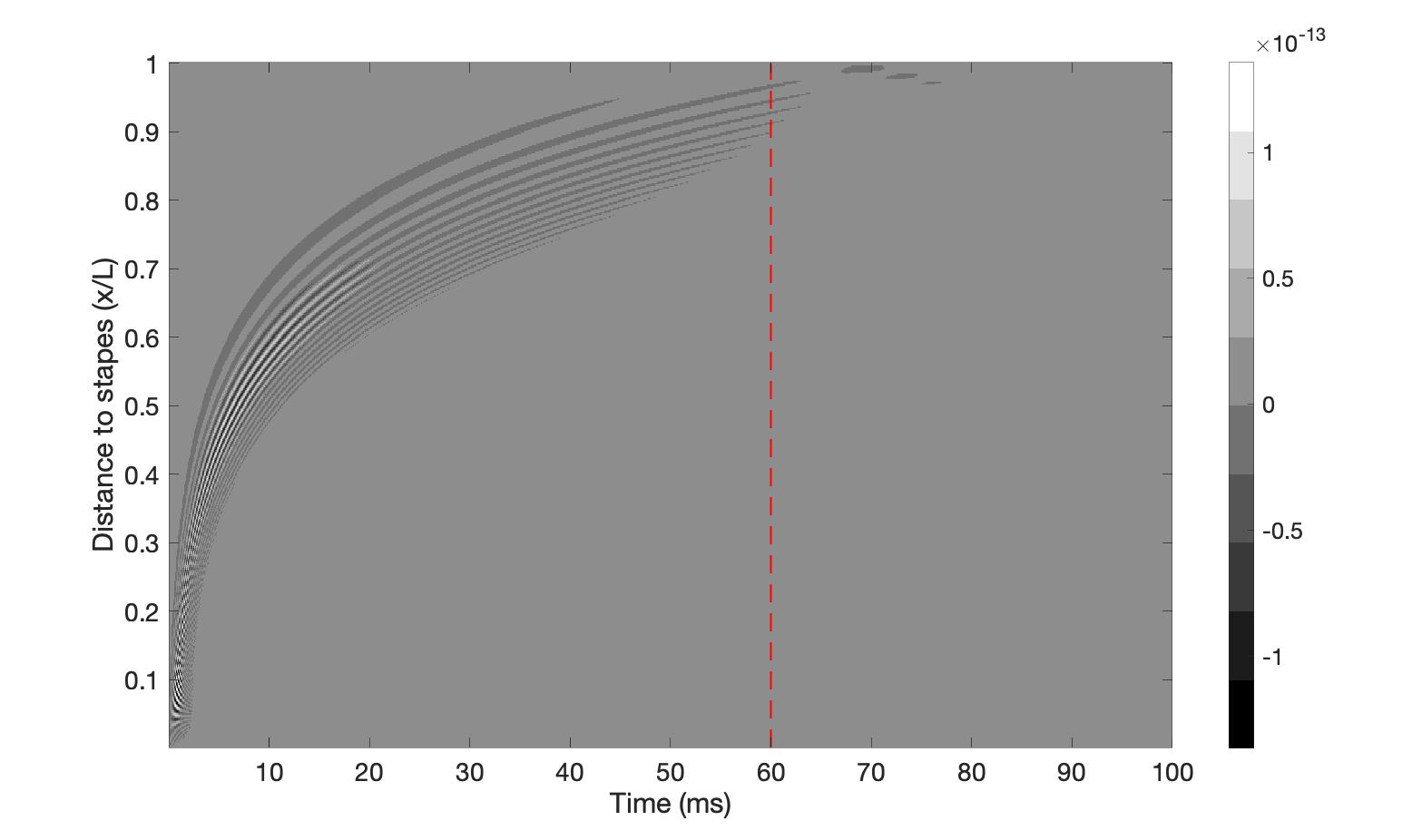}{9cm}{(b)}}
\caption{ \label{fig:3} Impulse travelling along cochlear in the (a) semi-discretised, and (b) proposed models. A transient time of around 60ms (red dash line) can be observed in both models.}
\end{figure*}
\subsection{Impulse response}
The displacement of each element of the BM can be viewed as the response of a band-pass filter with the centre frequency equal to the characteristic frequency corresponding to the position of that element along the BM. The impulse responses of these 'filters' were obtained by recording the response of the model to an impulse signal input at 0dB (SPL) over a duration of 100ms. The magnitude response of these 'filters' can then be inferred by transforming the impulse responses corresponding to each BM element to the frequency domain. The magnitude responses of five randomly selected elements ('filters') are plotted and compared to responses of the same elements in the semi-discretised model of \cite{elliott2007state} in Figure \ref{fig:2}. It can be observed that the filter responses in two systems show similar patterns, but interestingly the filters in the proposed model display a more sharp response with high suppression compared to those in the semi-discretised model.

Cochleagrams depicting the responses to a click (impulse) stimulus obtained from both the proposed jointly discretised system and the semi-discrete system from \cite{elliott2007state} are shown in Figure \ref{fig:3}. It also shows a similar travelling oscillation from base to apex for both systems, and the transit time for both is approximately 60ms.
\begin{figure*}[ht!]
\parskip=6pt
\baselineskip=12pt
\noindent
\figline{\fig{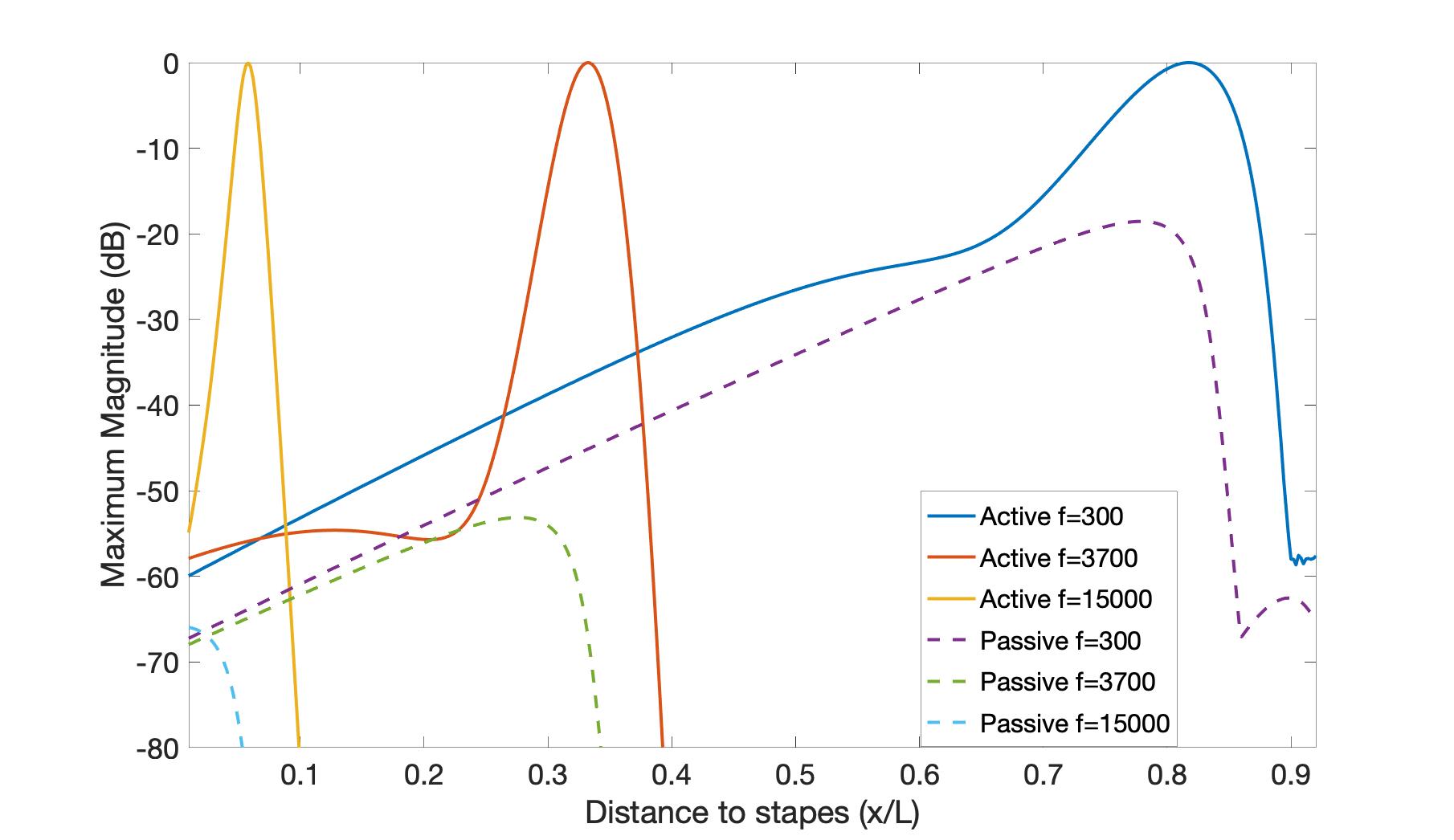}{.45\textwidth}{(a)}\label{fig:6a}
\fig{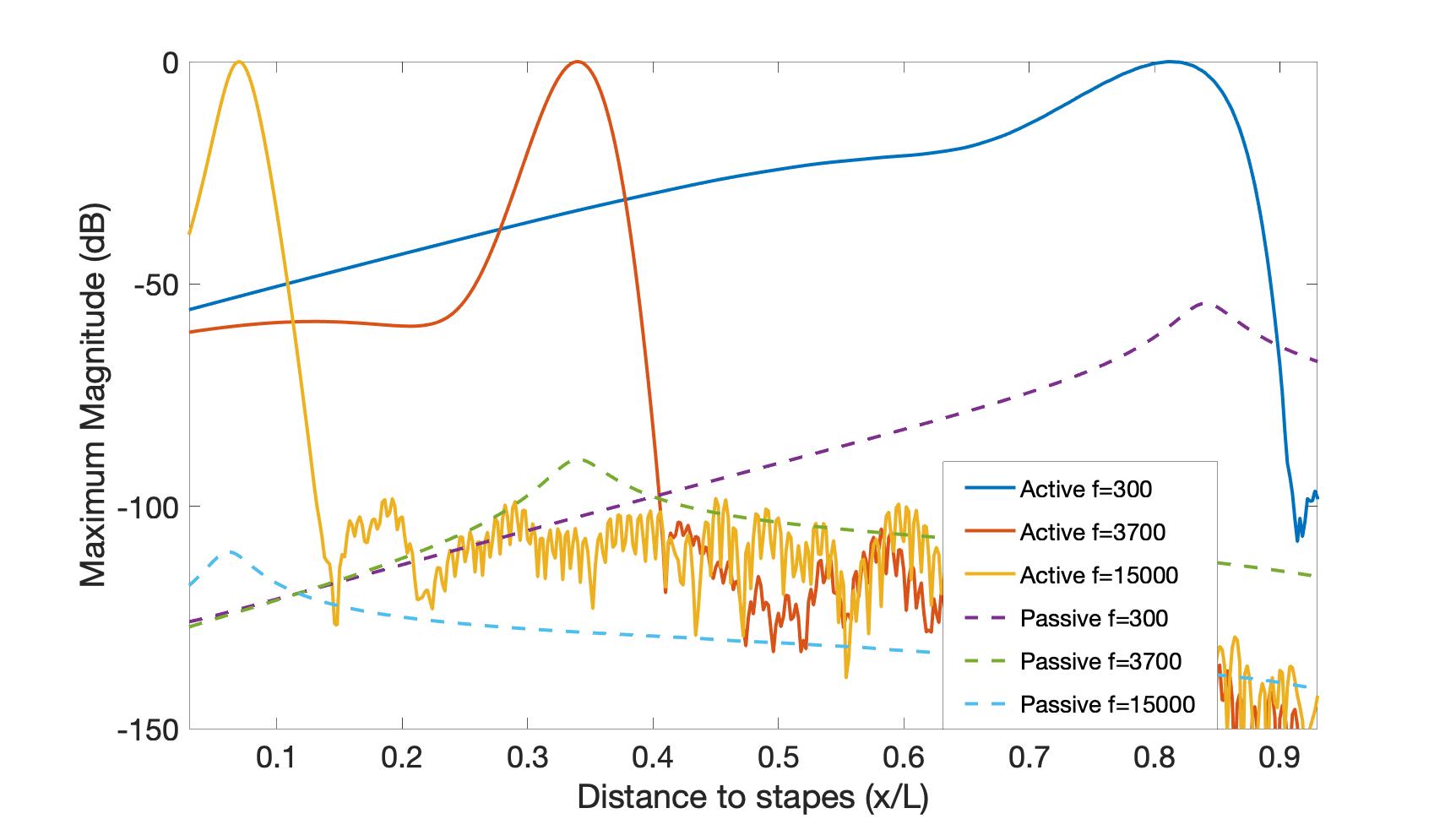}{.45\textwidth}{(b)}}\label{fig:6b}
\caption{ \label{fig:comp} BM response along its length to sinusoidal inputs with frequencies \textit{f}=300\textit{Hz}, 3700\textit{Hz} and 15000\textit{Hz} at 0dB (SPL) for the (a) proposed model; and (b) semi-discretised model. The characteristics of the proposed model are consistent  with  that  of  the  semi-discretised  model. }
\end{figure*}
\subsection{Single tone response}\label{sec:tone_resp}
\begin{figure*}[ht!]
\parskip=6pt
\baselineskip=12pt
\fig{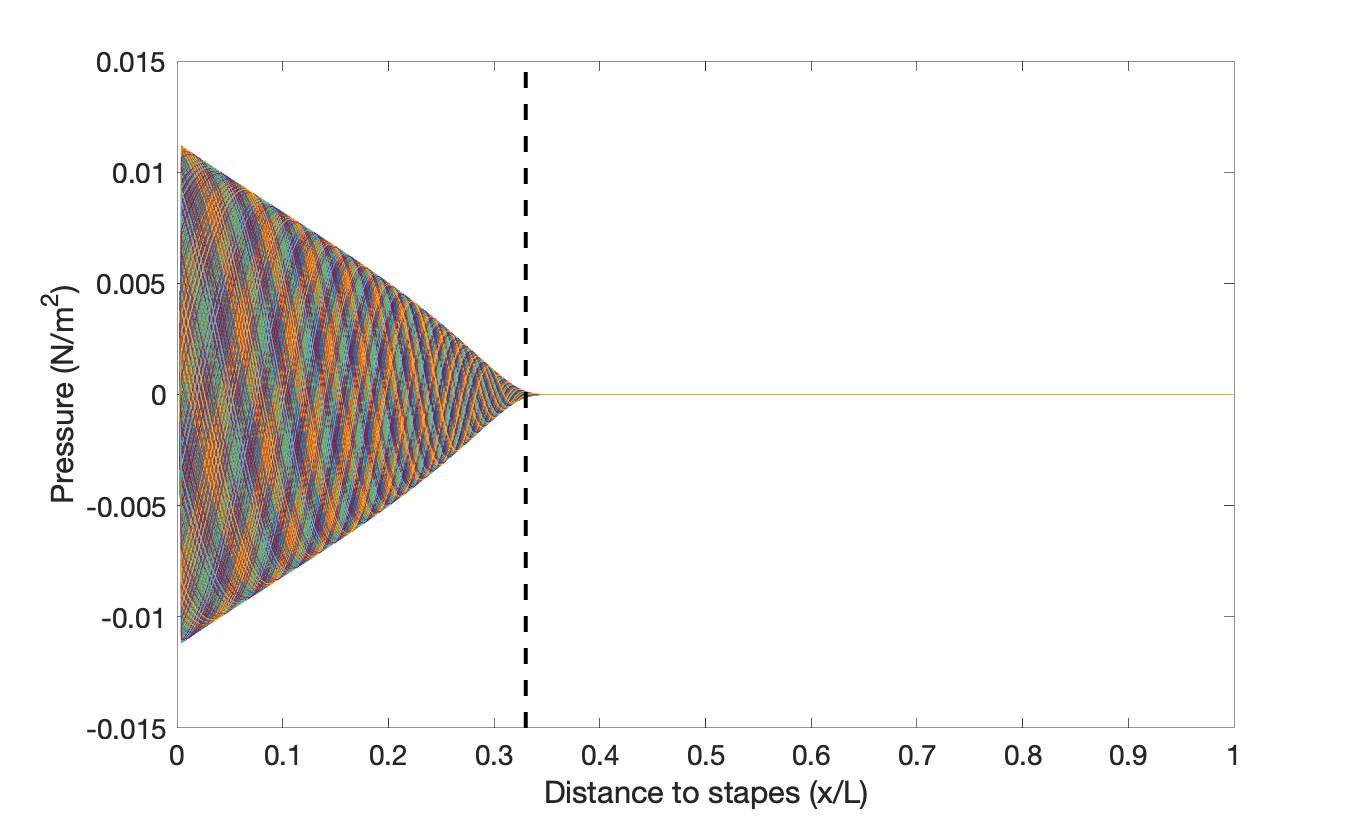}{.45\textwidth}{(a)}\label{fig:7a}
\fig{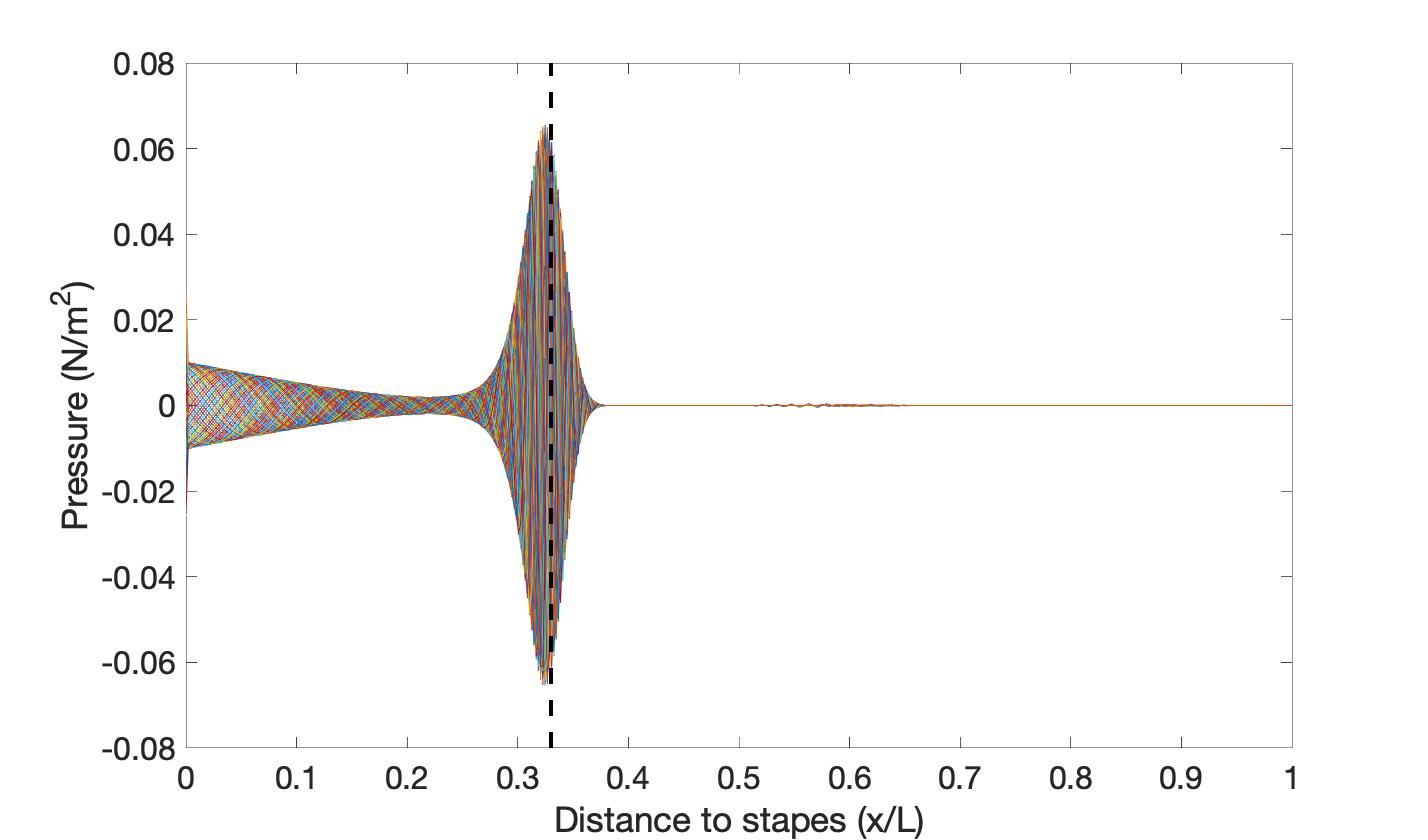}{.45\textwidth}{(b)}\label{fig:7b}
\caption{\label{fig:7} Pressure difference between upper and lower chambers for a sinusoidal input with \textit{f}=3700\textit{Hz} at 60dB (SPL) for jointly discretised (a) passive and (b) active models. The plots show pressure differences at all time steps superimposed on each other. In the passive model this pressure difference decreases steadily and goes to zero at the position of peak displacement. Whereas, in the active model, we can observe an additional peak in pressure difference at this position arising from the feedback.}
\end{figure*}

Given a single tone sinusoidal input, the BM element with characteristic frequency corresponding to the input frequency is expected to oscillate with the highest amplitude. To determine the frequency selectivity of the model, we plot the gain of each BM element filter at the input tone frequency. This is determined by computing the magnitude spectrum (via a Discrete Fourier Transform) of the displacement of each BM element in response to a sinusoidal input of 100ms duration under the same SPL of 0dB. The magnitude response is estimated from the last 30ms since the model has a transient time (time taken for signal to travel from base to apex of the cochlea) of around 60ms (refer Figure \ref{fig:3}). We plot this for three different input tones,  $f_1$=15000Hz, $f_2$=3000Hz, and $f_3$=300Hz, in Figure \ref{fig:comp} and compare it with the corresponding gains for the semi-discretised model described in \cite{elliott2007state} as well as passive versions (without active feedback) of both the jointly discretised and semi-discretised models.
It should be noted that the parameters for the passive model were slightly different from those of the active model, and details can be found in \hyperref[app:E]{Appendix E}. As seen from Figure \ref{fig:6a}, the oscillation positions in the BM for both active and passive models match the expected positions for input stimuli of different frequencies. The BM responses obtained from the active model were sharper compared with those from the passive model, and the increased gain and selectivity of the active models are also evident. These characteristics are also consistent with those of the semi-discretised model from \cite{elliott2007state}, which are shown in Figure \ref{fig:6b}.

\begin{figure*}[ht!]
\baselineskip=12pt
\fig{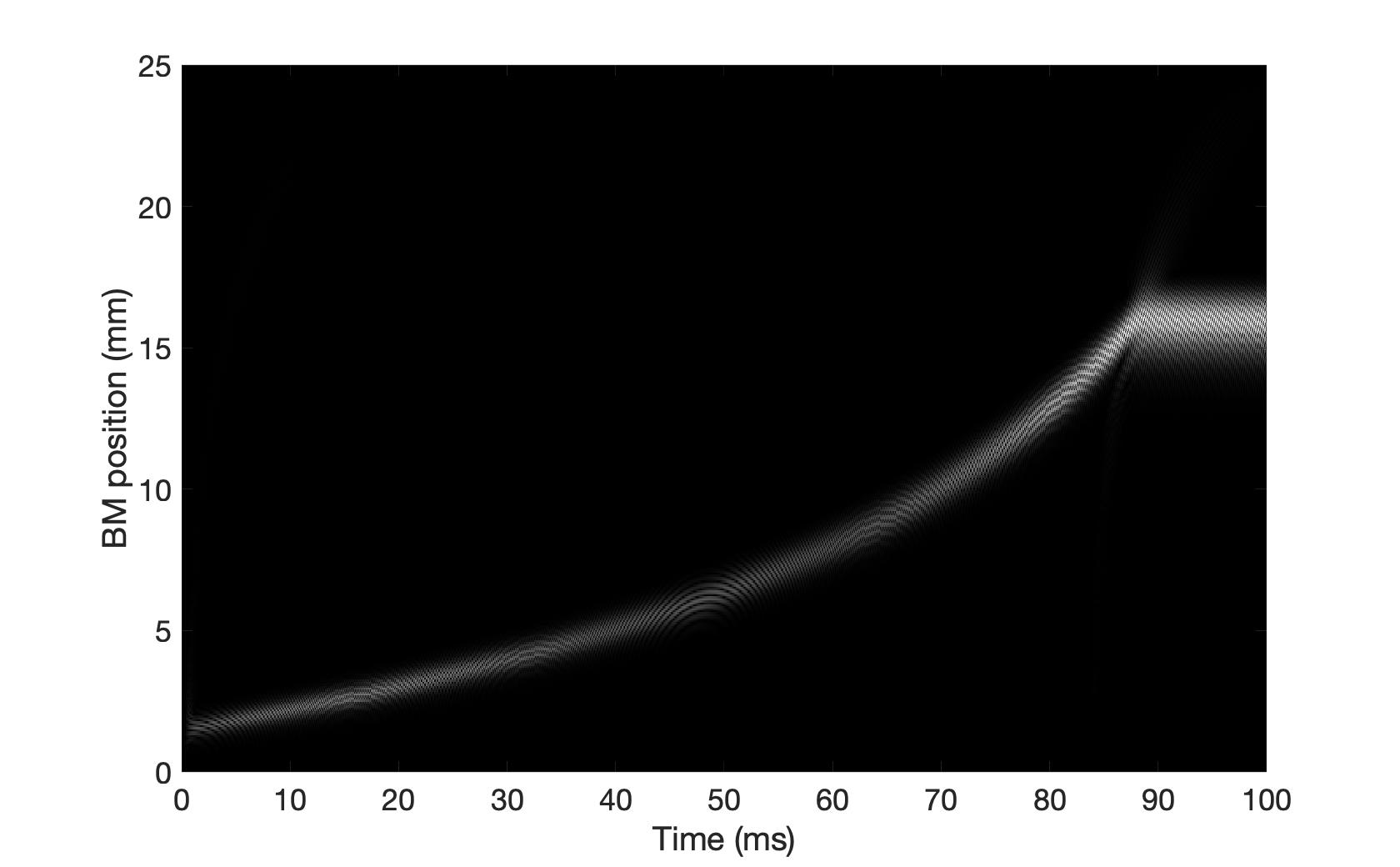}{.32\textwidth}{(a)}\label{fig:8aa}
\fig{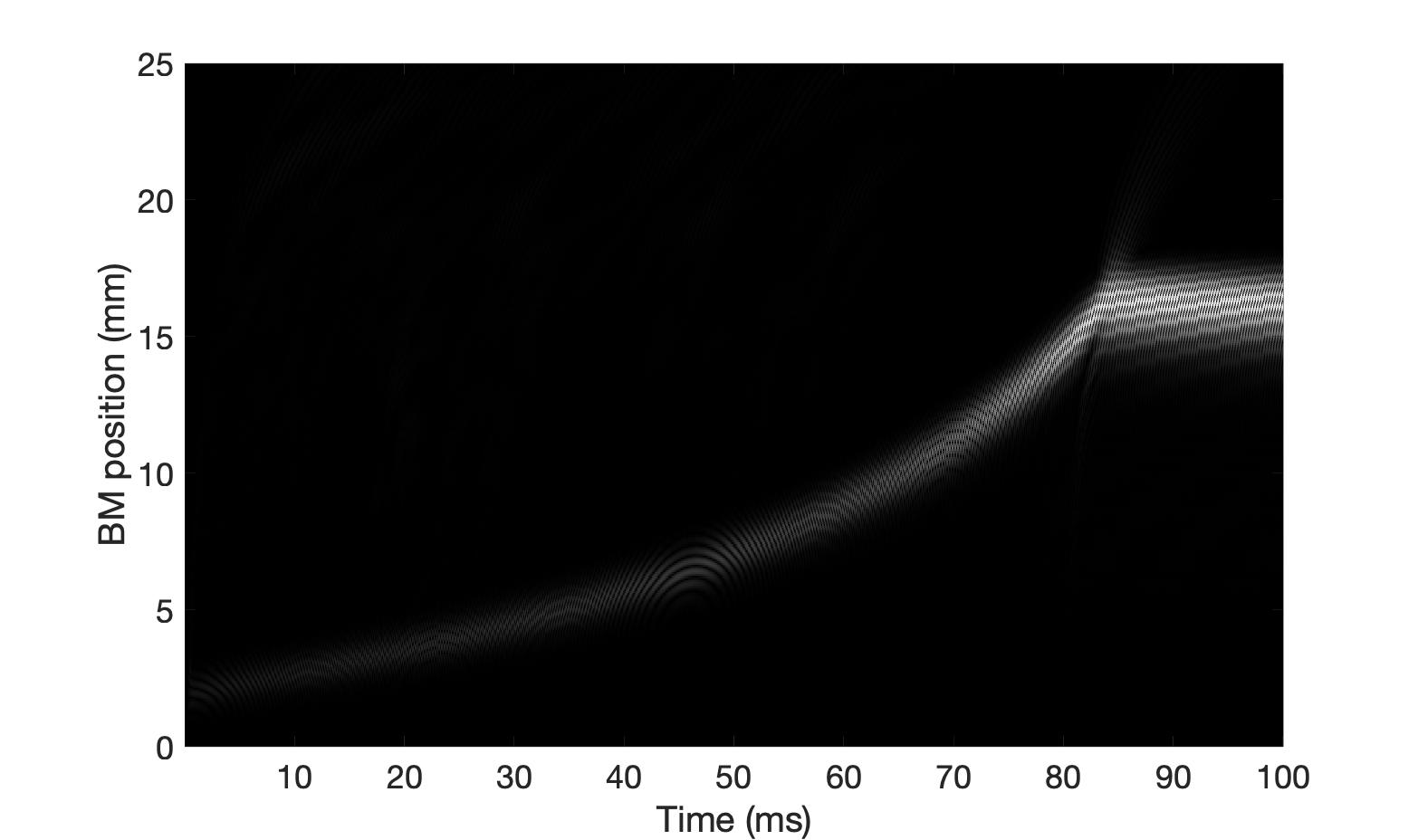}{.32\textwidth}{(b)}\label{fig:8bb}
\fig{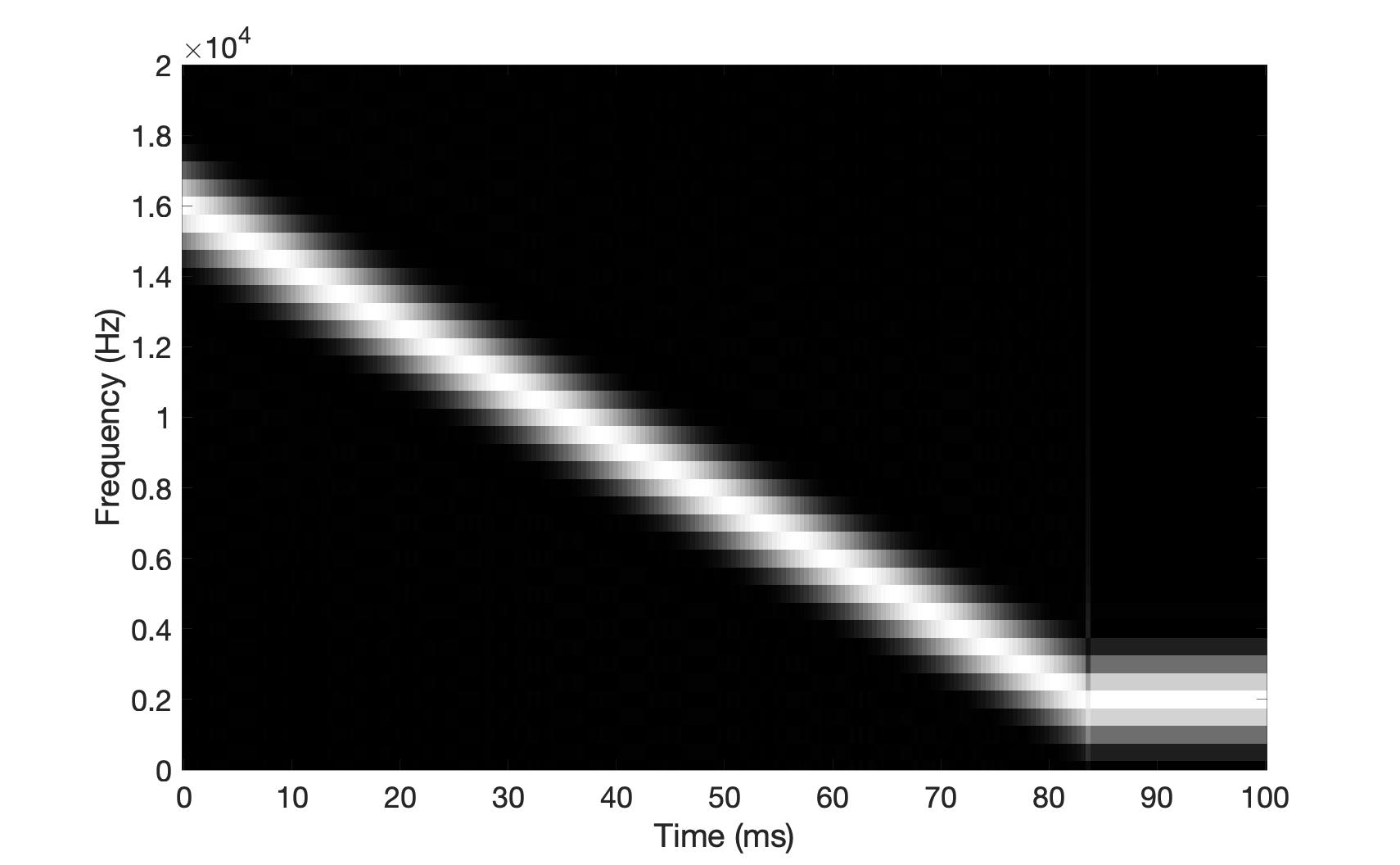}{.32\textwidth}{(c)}\label{fig:8cc}
\caption{\label{fig:99} Comparison of cochleagram and spectrogram given the same tme-varying signal of duration 100ms under SPL=0dB: (a) cochleagram of the proposed jointly discretised model; (b) cochleagram of the semi-discretised model; (c) spectrogram of the time-varying stimulus. The frequency is linearly varying from 16Khz to 2Khz. It is observed that the cochleagrams from both the proposed and  semi-discretised model show a similar pattern, and they also display similar information as the spectrogram, but with more detail in the low frequency regions.}
\end{figure*}

In addition to the BM displacement, the pressure across each BM element given a sinusoidal input with $f=3700$ Hz was computed for both passive and active models and are shown in Figure \ref{fig:7}. Once again, only the last 30ms of the response is shown, with the pressure at each time step within this duration superimposed over once another. In the passive model, as shown in Figure \ref{fig:7a}, the pressure differences decrease and disappear at the expected position for peak displacement. Whereas in the active model, in addition to the pressure component present in the passive model the active feedback component $p_{n}^a$, there is an additional peak in the pressure difference at the expected position for peak oscillation, as can be seen from Figure \ref{fig:7b}.

\subsection{Chirp response}
Finally, we also compare the chirp response of the proposed jointly discretised model with that of the semi-discretised model from \cite{elliott2007state}. Given a sinusoidal input with a time-varying frequency linearly decreasing from 16kHz to 2kHz over a duration of 100ms, cochleagrams from the model responses are compared with each other in Figure \ref{fig:99} (a spectrogram of the input signal is also provided for reference).

The cochleagram takes the absolute values of BM displacements as a function of both time and position (along the BM). Figures \ref{fig:8aa} and \ref{fig:8bb} show the cochleagrams from the proposed jointly discretised and semi-discretised models respectively. Note that the basal end responds to high frequencies and the apical end to low frequencies. From the two cochleagrams, it is clear that both models have similar spectral responses. In both cases, the positions of maximum BM response to the continuously changing input frequency (ranging from 16Khz at 1.5mm to 2kHz at 16.1mm along BM position) match closely. 

\begin{figure*}[ht!]
\parskip=6pt
\baselineskip=12pt
\noindent
\figline{\fig{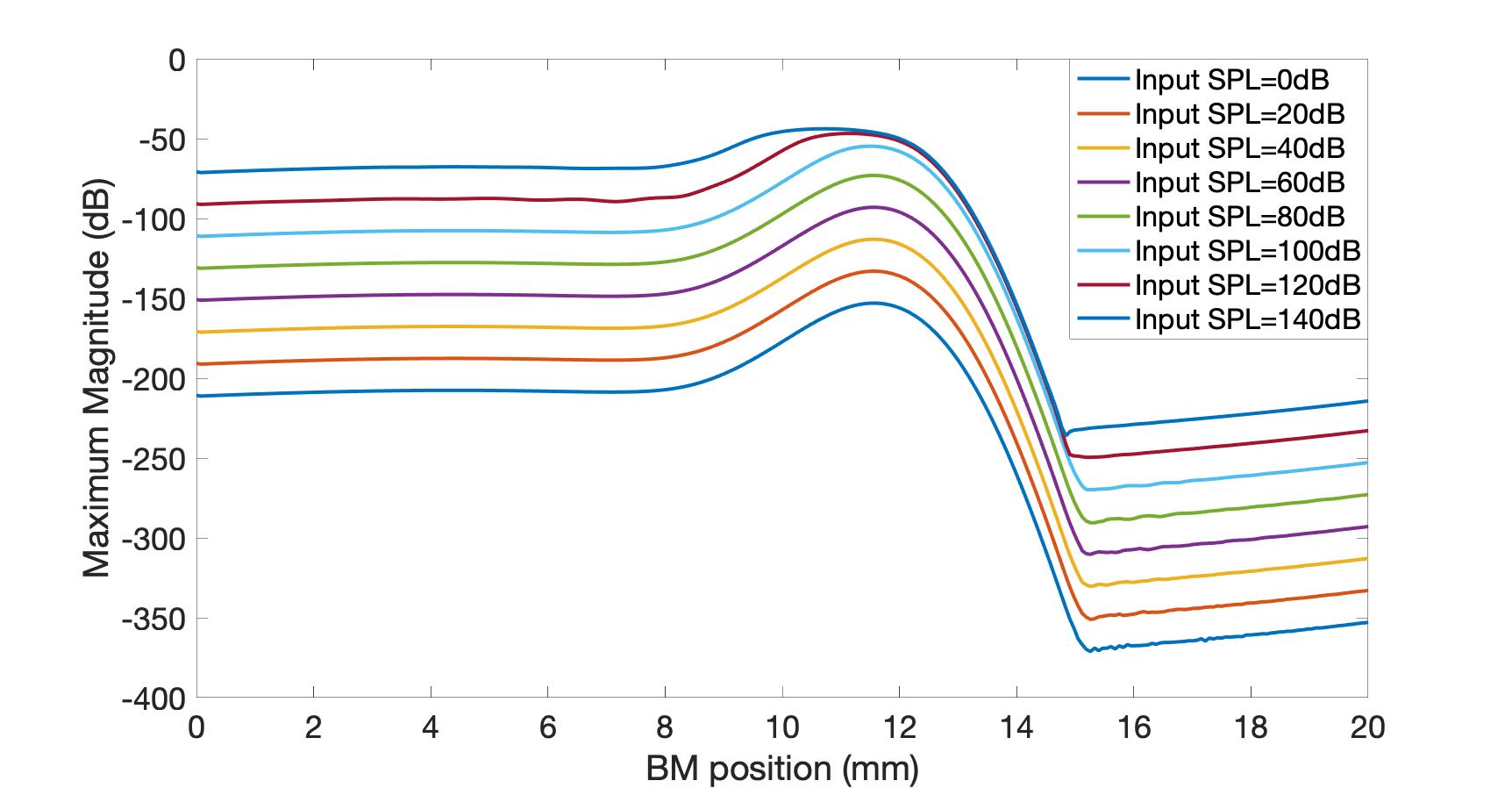}{8cm}{(a)}\label{fig:9a}
\fig{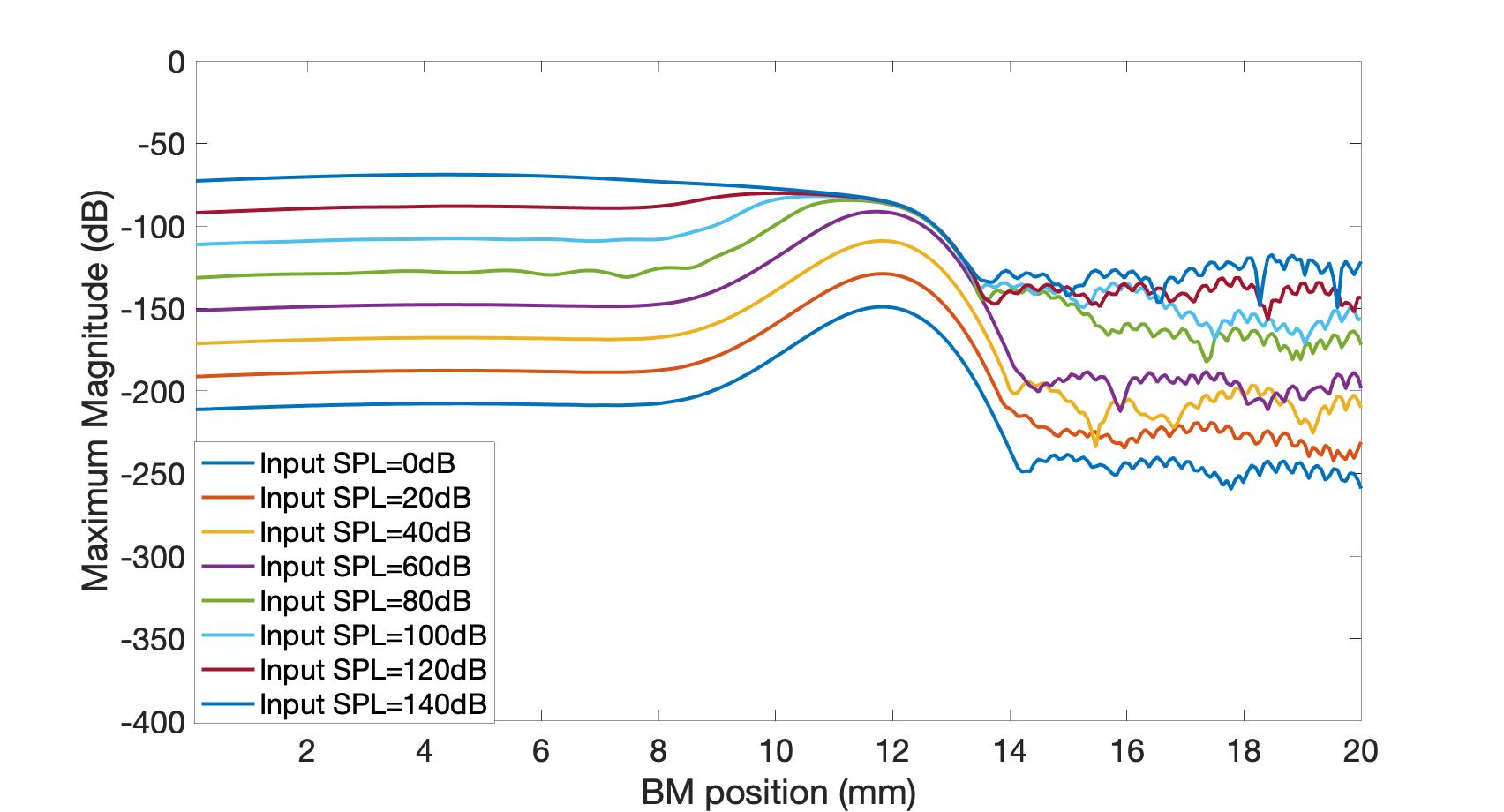}{8cm}{(b)}}\label{fig:9b}
\caption{ \label{fig:77} BM response along its length to sinusoidal inputs with a frequency of 3700Hz and various sound pressure levels ranging from 0dB (SPL) to 140dB (SPL) in steps of 20dB: (a) proposed jointly discretised model; (b) semi-discretised model.  The unequal spacing between the BM responses reveals the dynamic range compression in the model. The extent of the dynamic compression can be inferred by taking the ratio of the increase in magnitude of the BM response at 12mm (corresponding to 3700Hz) to the increase in input signal magnitude as it goes from 80dB(SPL) to 140dB(SPL).}
\end{figure*}

\section{Model Characteristics}\label{sec:5}
In addition to spectral decomposition of the input signal based on the tonotopy of the BM that was the focus of section \ref{sec:4}, there are two other characteristics of a cochlear model that are of great interest. Firstly, the dynamic range compression that is expected to arise from the effect of the nonlinear active feedback in the system; and secondly the stability of the model and in particular the choice of discretisation time step that leads to a stable model. Both of these are the focus of this section.

\subsection{Dynamic range compression}
Dynamic range compression in a cochlea refers to its ability to compress sound inputs across a large dynamic range into neural representations with a much smaller dynamic range. In the context of cochlear models, this can be analysed by quantifying the dynamic range of the BM displacements and comparing to the dynamic range of the input. In this section we describe the analyses carried out to validate that the dynamic range compression in the proposed jointly discretised model matches that of the semi-discretised model in its ability to emulate this characteristic of human cochlea. To do this, the BM displacements in response to a single tone input at different input sound pressure levels in the range of 0dB to 140dB, with steps of 20dB, were obtained. The nonlinear feedback in the model should lead to higher gain for input signals at low SPL and lower gain for input signals at high SPL, leading to dynamic range compression. The BM responses (refer to section \ref{sec:tone_resp}) to sinusoidal input signals with a frequency of 3700Hz at different input SPLs are shown in Figure \ref{fig:77}.

It can seen in Figure \ref{fig:9a} that at low to moderate input SPLs, the gains are relatively constant. For inputs ranging from 0db(SPL) to 80db(SPL), the BM response at the position corresponding to 3700Hz ($\sim$12mm) also has a dynamic range of around 80dB. However, for inputs from 80dB(SPL) to 120db(SPL), the dynamic range of the corresponding BM responses is lower than 40dB and finally the BM responses to input at 140dB(SPL) and 120dB(SPL) have more or less identical amplitudes. This pattern of dynamic range compression is similar to that observed in the semi-discretised model of \cite{elliott2007state} which is shown in Figure \ref{fig:9b}.

\begin{figure}[ht!]
    \centering
    \includegraphics[width=0.5\textwidth,height = 5cm]{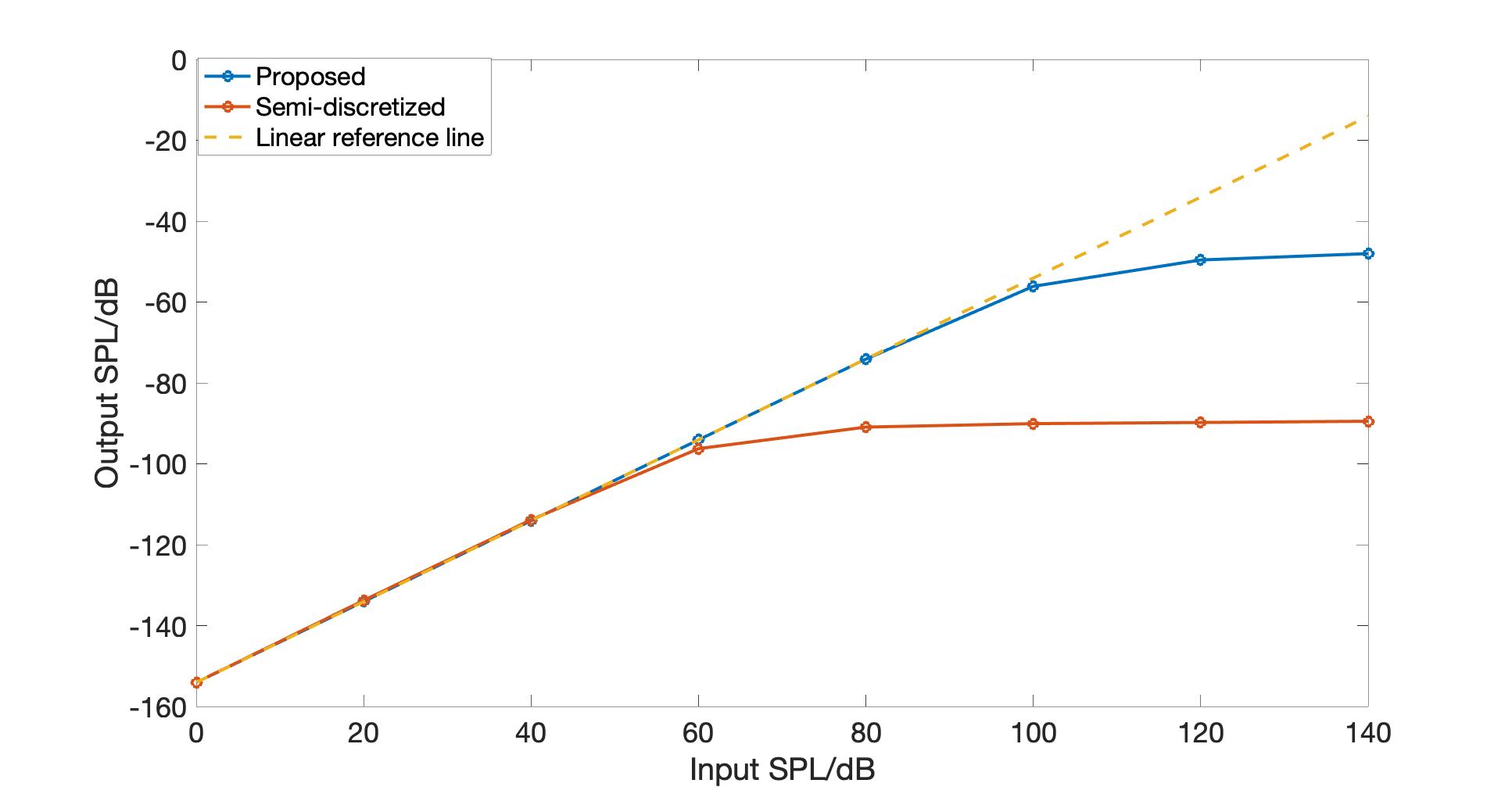}
    \caption{Input SPL vs BM response power at position of maximum response given a single tone input of frequency 3700Hz. The red dash line is a linear reference line. Both the proposed and semi-discritised model shows dynamic range compression.}
    \label{fig:88}
\end{figure}

Finally, Figure \ref{fig:88} shows a plot of input SPL (in dB) against output power (in dB), at the BM position of maximum response. From the plot it is clear that the relationship is linear for input SPLs from 0dB to 80dB. For input SPLs between 80dB and 100dB, the corresponding change in output power is 18dB (from -74dB to -56dB) and the output saturates rapidly after that point. This is similar to the input-output trend for the semi-discretised model with the difference being that the jointly discretiesd model starts to saturate at an input level of around 80dB(SPL), while the semi-discretised model saturates at an input level of around 60dB(SPL). Similar characteristics were also observed for the TM responses in the model.

\noindent

\subsection{System stability}\label{sec:stability_analyses}
In section \ref{sec:3.4}, we discussed the system stability for the linear active cochlear system that is jointly discretised in both spatial and temporal domains, where the magnitude of the largest eigenvalues should be smaller than 1. Table \ref{tab:22} displays the magnitude of the largest eigenvalues of matrix $\mathbf{E}$ for a linear active cochlear model given different sampling frequency $fs$ within the range of [48Khz 192kHz] with a step size of 16kHz. It is observed that the magnitude of the largest eigenvalues is larger than one with the sampling frequency smaller than 112Khz, indicating an unstable system. In order to guarantee the system is never unstable, the sampling frequency was chosen to be $fs=128$kHz with a relatively small time step size of 7.8 microseconds. All other experimental settings are same as in Table \ref{tab:real1}. 

\begin{table}[ht]
\caption{Magnitude of maximum eigenvalue for systems discretized with different sampling frequency $fs$}
\begin{adjustbox}{width=0.45\textwidth}
\begin{tabular}{| c|c|c |c |c |c |c | c |c|c|c|}
\hline
fs (kHz) & 48 &64 & 80 &96 & 112 & 128 & 144 & 160 & 176 & 192 \\
     \hline
     Magnitude & 1.89 &1.23 &1.07 & 1.02&1.0 & 1.0 & 1.0 & 1.0 & 1.0 & 1.0 \\ 
     \hline
     Stability &
     \multicolumn{4}{c|}{Unstable} & \multicolumn{6}{c|}{Stable} \\
     \hline
\end{tabular}
\end{adjustbox}
\label{tab:22}
\end{table}

Similarly the nonlinear feedback was analysed by computing the eigenvalues of each time step, as $\mathbf{E}$ is time varying. It was found that all the eigenvalues across time were all within the unit circle, which indicates that the nonlinear active model is also stable.
\begin{figure*}[ht!]
    \centering
    \includegraphics[width=1\textwidth]{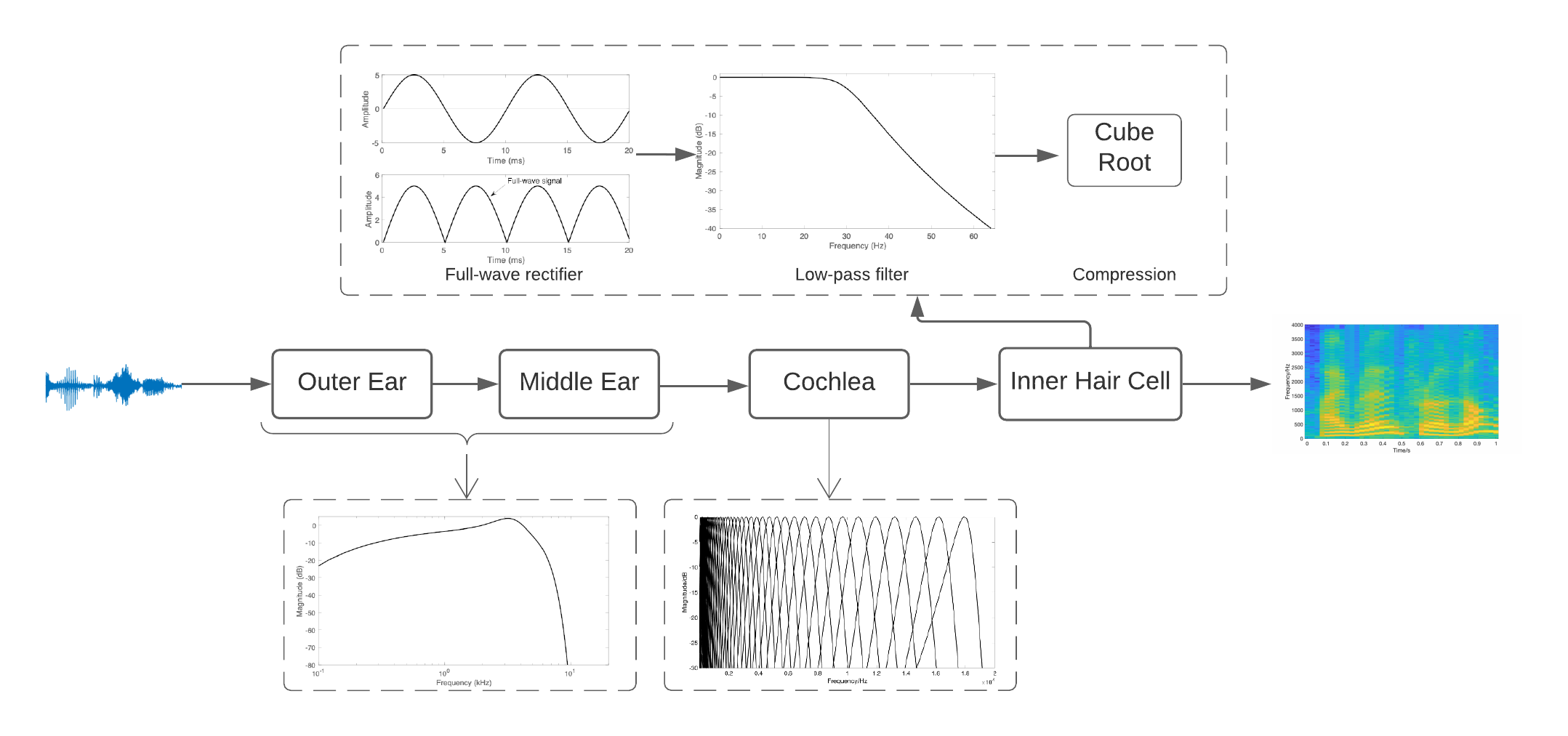}
    \caption{System block diagram for complete human auditory processing, including functions of outer and middle ear, proposed cochlear model, and inner hair cells. The characteristics are displayed in terms of frequency responses under each block.}
    \label{fig:18}
\end{figure*}
\section{Speech}\label{sec:7}
As mentioned in the introduction, a key motivation underpinning the development of the proposed jointly discretised cochlea model is its potential application as a robust front-end for speech processing systems. The temporal discretisation allows for constant time step processing that is compatible with standard DSP systems, albeit at a high sampling rate, and the model formulation entirely in terms matrix operations makes its integration with deep learning systems and implementation on GPUs a promising avenue for development.

We combine the proposed model of the cochlea with discrete-time models of the outer ear, middle ear and inner hair cells to simulate auditory processing in the human ear for cochleagram estimation. In this section we present a comparison of such cochleagrams with conventional spectrograms for speech inputs under different sound pressure levels.
It is expected that the cochleagram i) captures more detailed information in low frequency regions where most of the speech content lies since the filter design in the cochlear model matches human auditory perception
and more importantly ii) shows greater invariance given the same input with different amplitudes compared with spectrograms. This is expected due to the dynamic compression of the cochlear model, and would make the system more robust to input SPL variations. 

\subsection{Human auditory system structure}
A block diagram showing the different elements of the human auditory system is shown in Figure \ref{fig:18}. As shown, sound inputs are first processed through the outer ear and the middle ear, which effectively amplify the 3kHz to 5kHz band of the input signal. The middle ear and outer ear are modelled as per the transfer function proposed by Terhardt \cite{terhardt1979calculating}:

\begin{equation}\label{eq:40}
    A(f) = -3.64f^{-0.8}+6.5e^{(-0.6(f-3.3)^2)}-10^{-3}f^4
\end{equation}
where $f$ is the characteristic frequency of each BM element, and $A(f)$ is the magnitude response at frequency $f$. This magnitude response of the combined outer and middle ear is a bandpass filter, as shown in Figure \ref{fig:18}. 

The output from the middle ear is then the input to the proposed jointly discretised cochlear model, and this is followed by an inner hair cell (IHC) model. The inner hair cells convert BM displacement into electrical impulses and pass them on to the brain via the auditory nerve. This action is typically modelled as a full-wave rectifier followed by a low-pass filter and cuberoot compression, which together act as an envelope detector, prior to converting them to a sequence of electrical impulses. In this work we do not convert the signals to a sequence of impulses but all other elements of the inner hair cell model are implemented. The low-pass filter element of this IHC model is a simple 1st order filter with a transfer function given as:
\begin{equation}\label{eq:41}
    H(z) = \frac{1-c_0}{1-c_0z^{-1}}
\end{equation}
with,
\begin{equation}\label{eq:42}
    c_0=e^{\frac{-30*2\pi}{fs}}
\end{equation}
where $f_s$ is the sample frequency of the input stimulus. This value of $c_0$ leads to a low-pass filter with a cut-off frequency of 30Hz. Finally, the output from the low-pass filter undergoes power-law compression ($y = x^{1/3}$).

\begin{figure*}[ht!]
\parskip=1pt
\baselineskip=1pt
\noindent
\figline{\fig{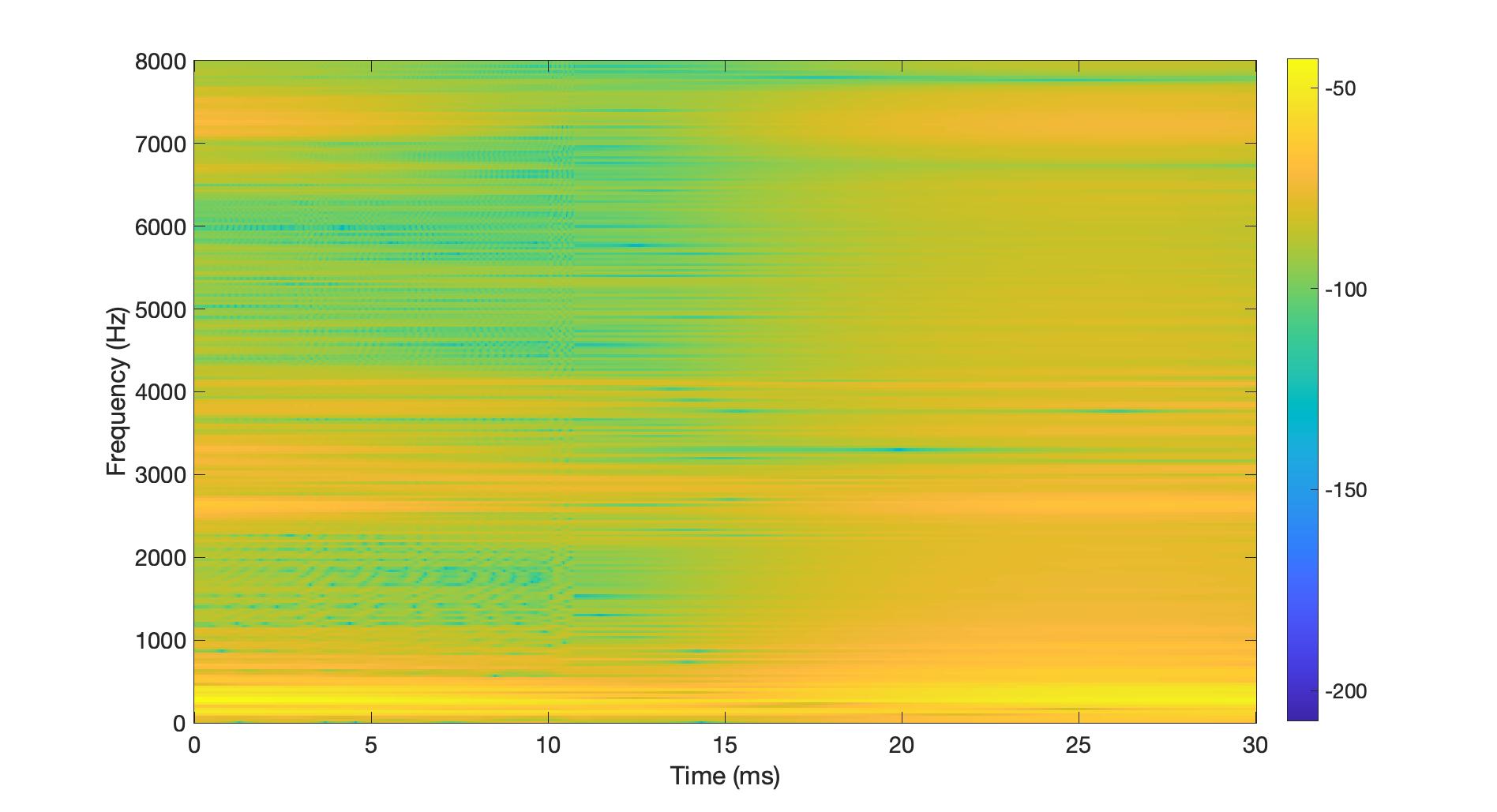}{.45\textwidth}{(a) Spectrogram of vowel 'iy': sample 1 from speaker 1}\label{fig:1sub}
\fig{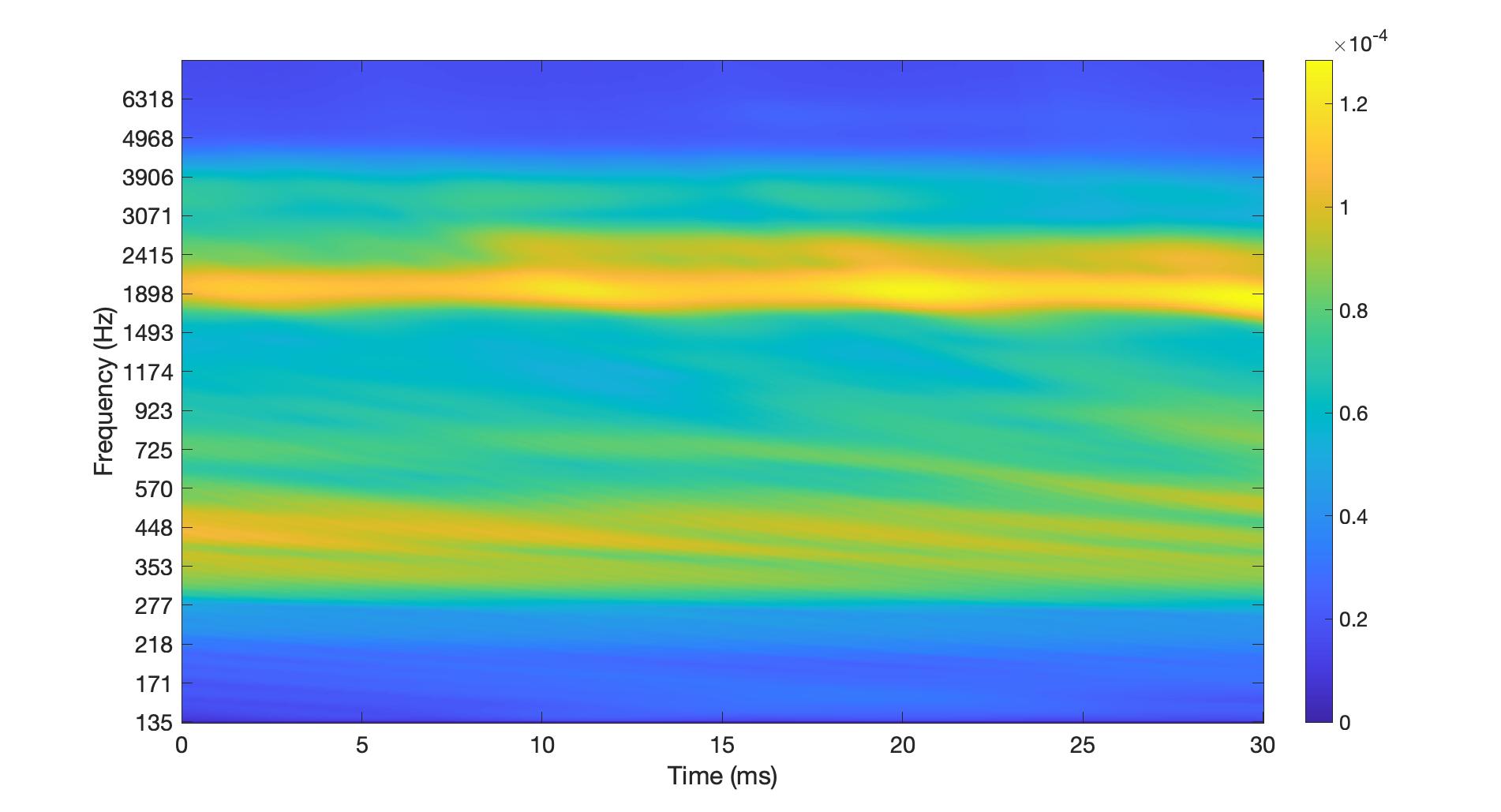}{.45\textwidth}{(b) Cochleagram of vowel 'iy': sample 1 from speaker 1}}\label{fig:2sub}
\figline{\fig{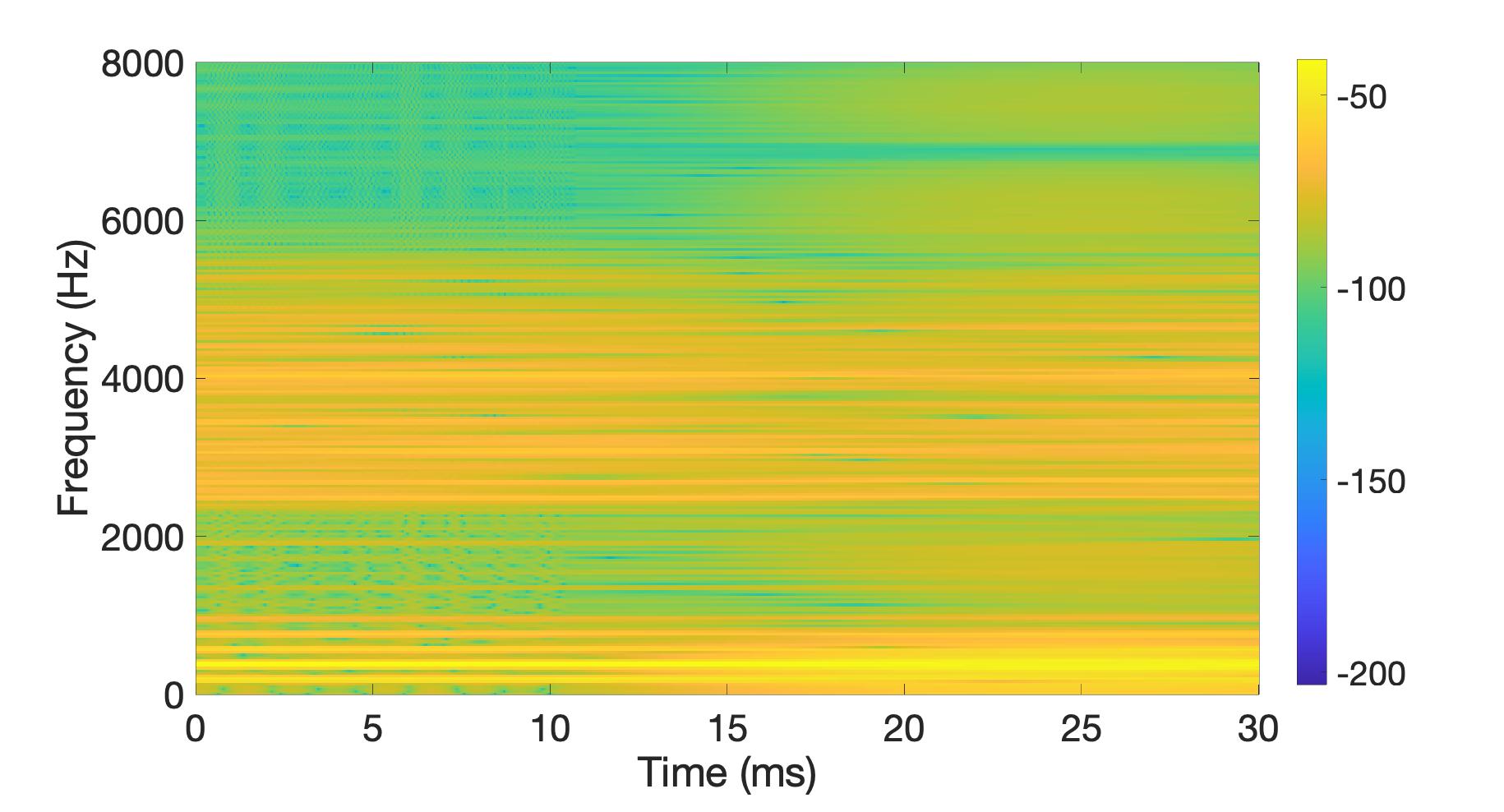}{.45\textwidth}{(c) Spectrogram of vowel 'iy': sample 2 from speaker 1}\label{fig:3sub}
\fig{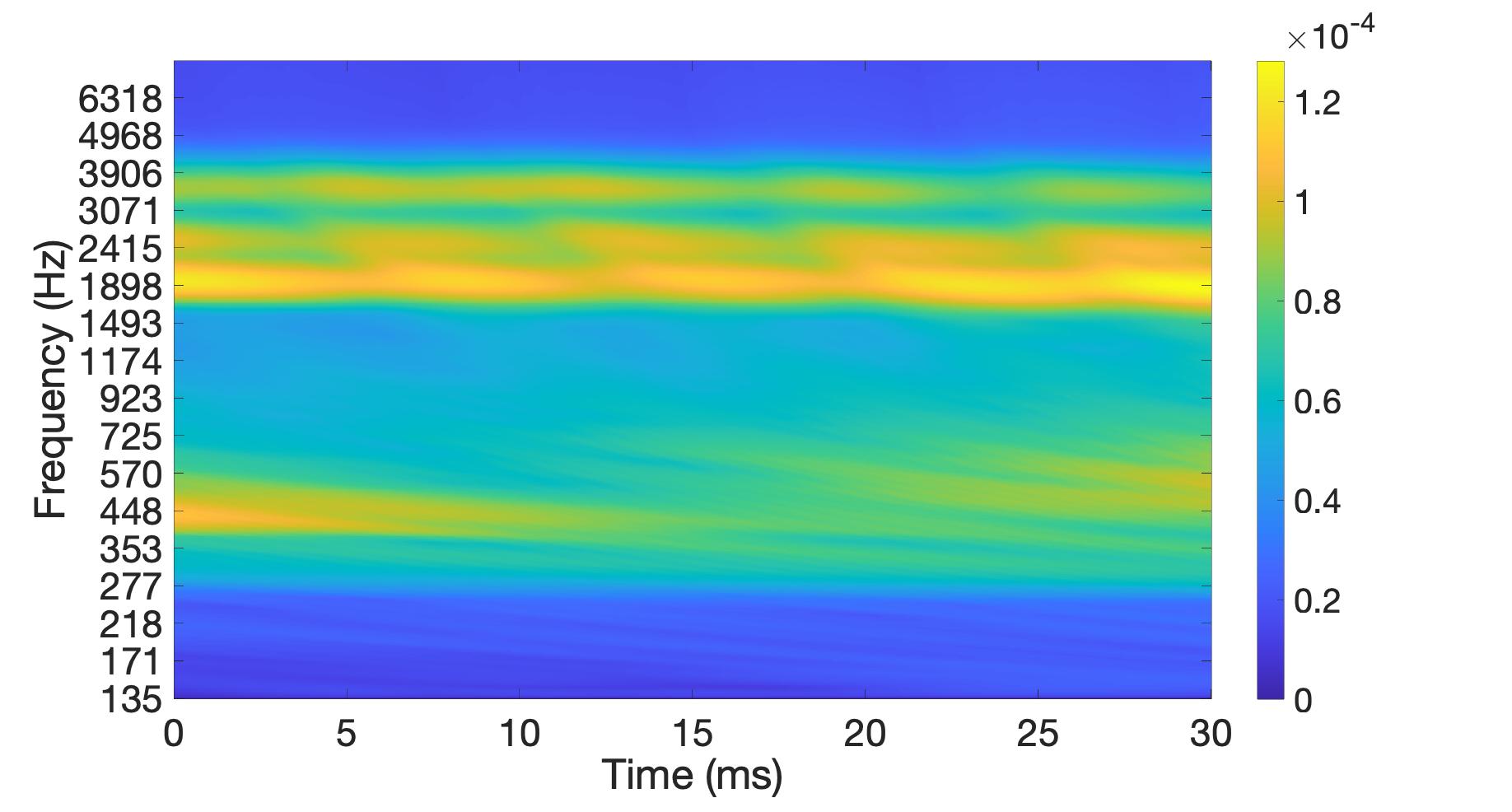}{.45\textwidth}{(d) Cochleagram of vowel 'iy': sample 2 from speaker 1;}}\label{fig:4sub}
\figline{\fig{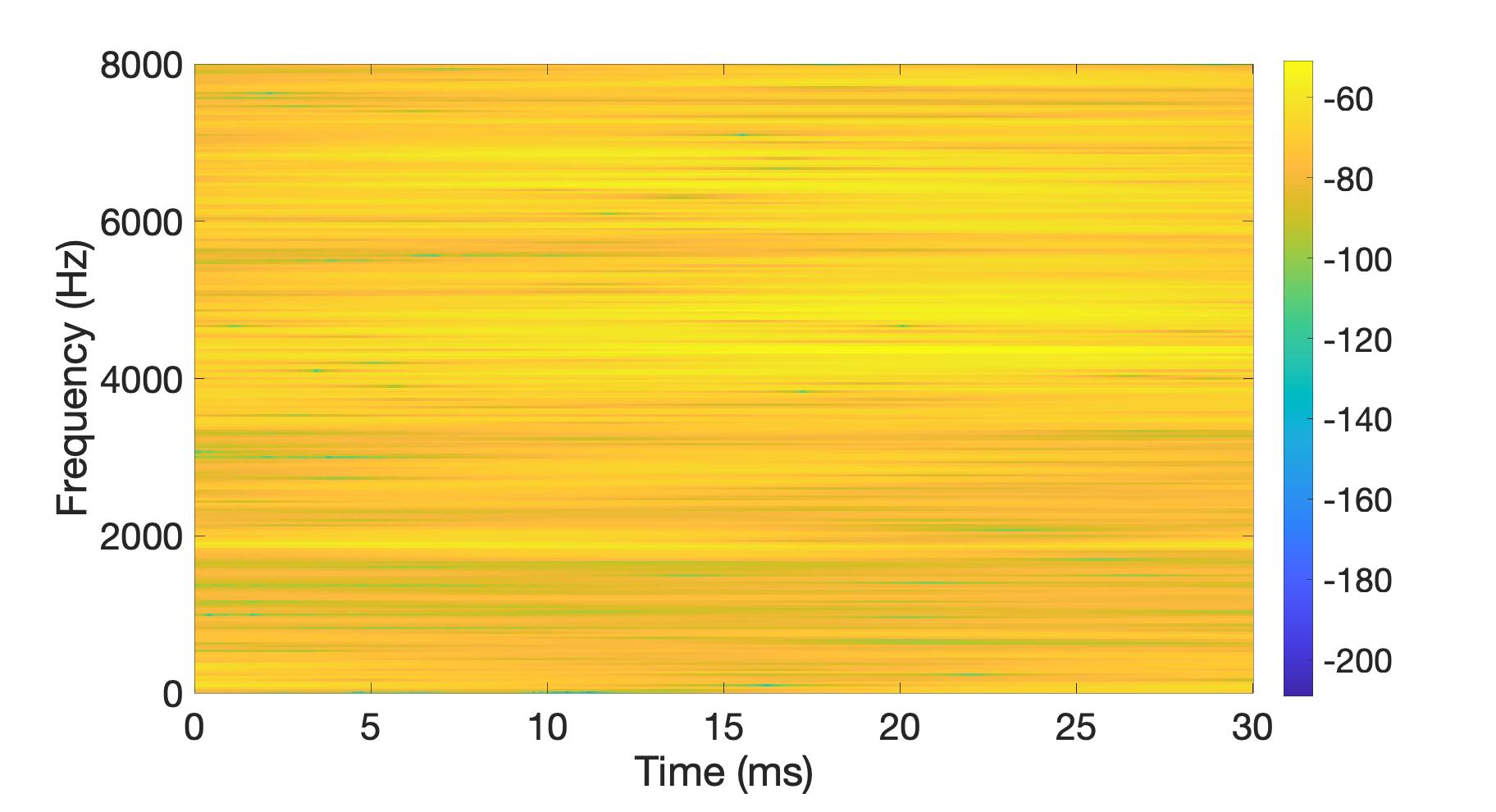}{.45\textwidth}{(e) Spectrogram of vowel 'iy': sample 1 from speaker 2}\label{fig:5sub}
\fig{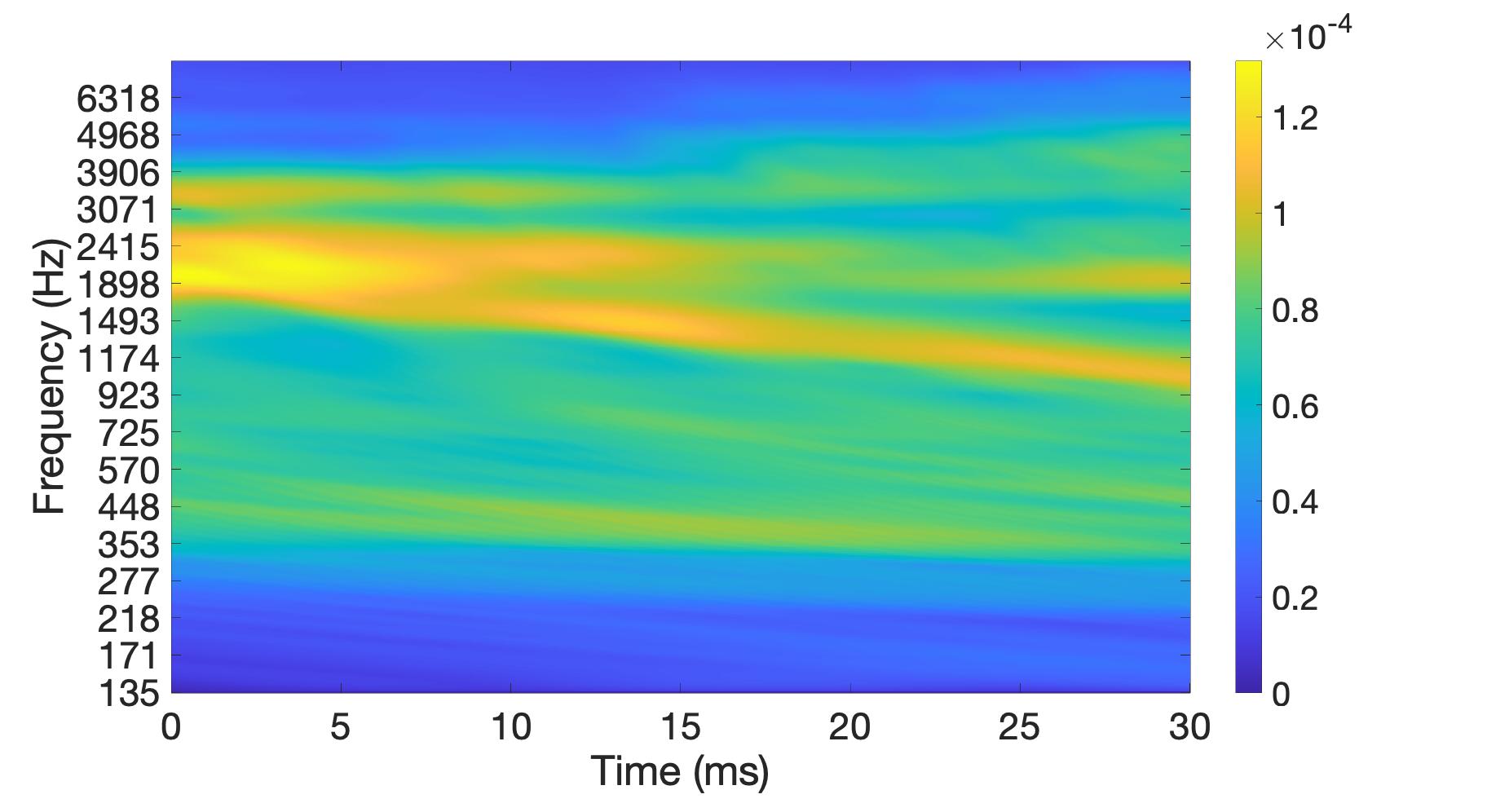}{.45\textwidth}{(f) Cochleagram of vowel 'iy': sample 1 from speaker 2}}\label{fig:6sub}
\figline{\fig{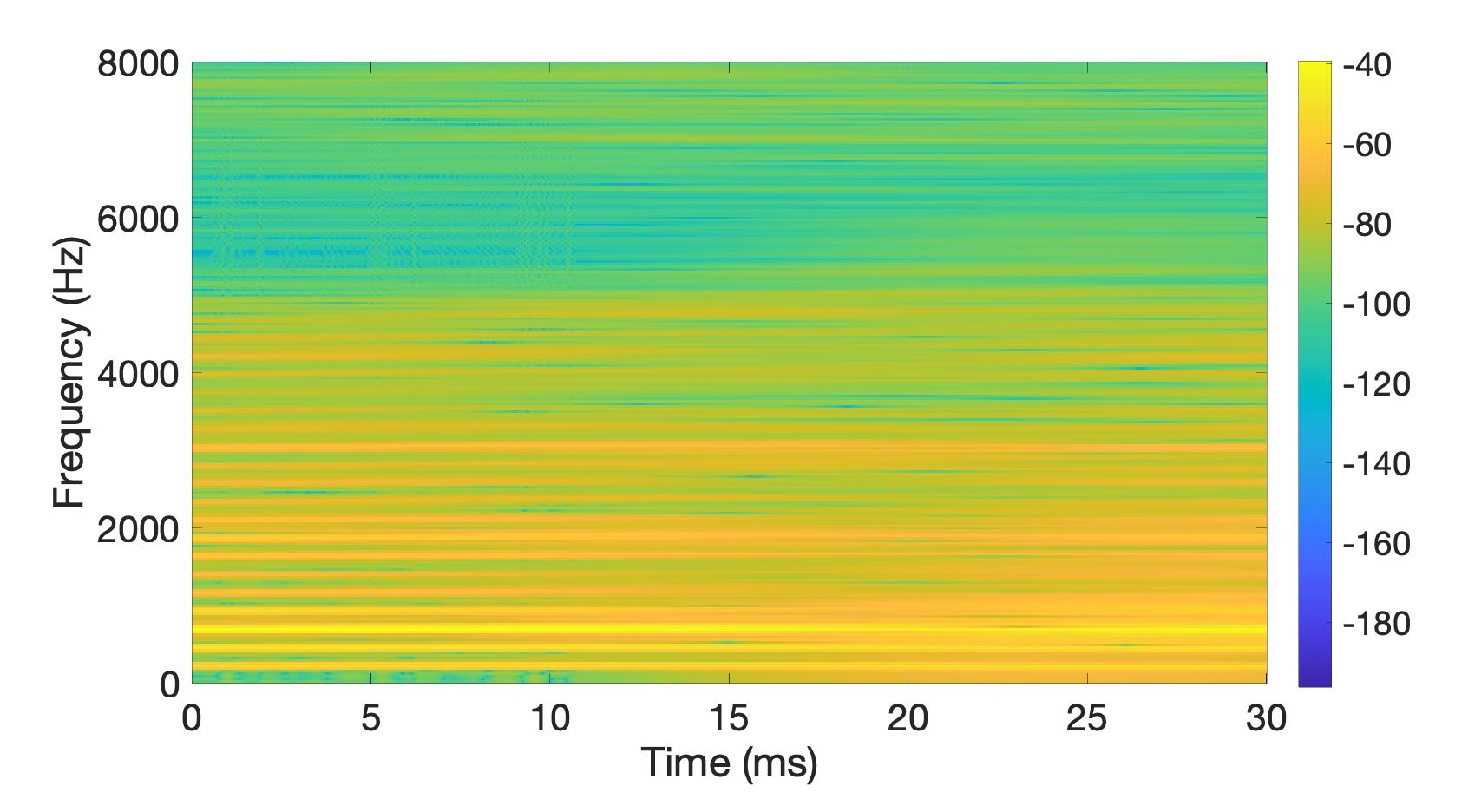}{.45\textwidth}{(g) Spectrgram of vowel 'ae': sample 1 from speaker 3}\label{fig:7sub}
\fig{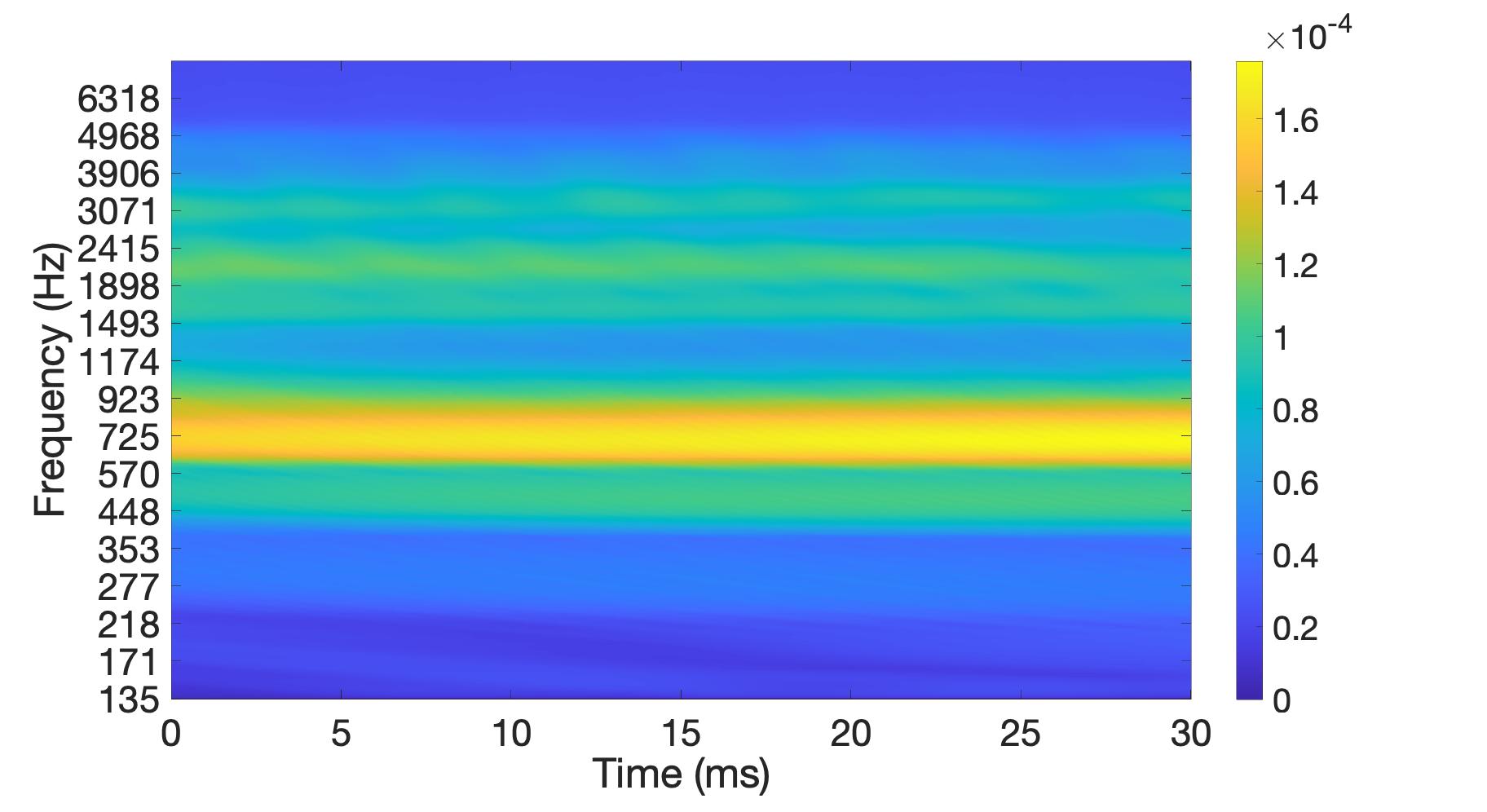}{.45\textwidth}{(h) Cochleagram of vowel 'ae': sample 1 from speaker 3}}\label{fig:8sub}
\caption{ \label{fig:cp11} Comparison of spectrograms
to cochleagrams of different vowels and different speakers with input at 0dB(SPL). Spectrograms are shown in dB and cochleagrams are shown as obtained after cuberoot compression. Cochleagrams can capture the same frequency components as spectrograms, but they exhibit greater visual similarity between same vowels and greater visual dissimilarity between different vowels.}
\end{figure*}

\subsection{Experimental settings}
The TIMIT database was utilised for all the experiments reported in this paper. Specifically, 1000 segments each from 20 vowels of the training partition of TIMIT \cite{zue1990speech} were used for the different comparison of cochleagrams generated from the proposed model with spectrograms.
To determine if the output of the proposed model are robust to input signal level variations, each vowel speech clip was scaled to simulate 7 different sound pressure levels ranging from 0dB to 120dB with a step size of 20dB. This was achieved by scaling the original signal, $x(n)$ as per:

\begin{equation}\label{eq:43}
    SPL = 20\mathrm{log_{10}}^{\mathrm{max}(|x(n)|)}+96
\end{equation}
 with the scaling factor, $s$, computed as:
\begin{equation}\label{eq:44}
    s = \frac{10^{(SPL-96)/20}}{(\mathrm{max}(|x(n)|)/\sqrt{2})}
\end{equation}

Prior to processing, the speech samples from TIMIT were upsampled from 16kHz to 128kHz, the sampling rate at which the proposed cochlear model is suggested to operate at. Spectrograms were computed with a window size of 10ms and to match the time resolution of the cochleagrams, the spectrograms were computed with overlapped windows with only one sample shift between consecutive windows.

\subsection{Comparisons between cochleagrams and spectrograms}
The spectrograms and cochleagrams of the same vowel and different vowels are compared in Figure \ref{fig:cp11}.  Spectrograms are represented in dB scale to enhance the smoothness and cochleagrams are shown in magnitude as they are already smoothed representations owing to the inner ear functioning as a low pass filter and the following cube root compression. Only information under 8kHz is shown as the original sample frequency in TIMIT was 16kHz. It is observed that spectrograms and cochleagrams in Figures \ref{fig:cp11}(a) and \ref{fig:cp11}(b) shows a similar pattern for vowel 'iy', which all displays the frequency components at frequency $\textit{f}_1$ $\approx$ 450Hz, and $\textit{f}_2$ $\approx$ 2.5kHz. It is obvious to observe more detailed information in the low frequency region in the cochleagram, especially regarding the fundamental frequency $f \approx 300$Hz compared with the corresponding spectrogram. More importantly, the cochleagrams for the vowel 'iy' spoken by the same speaker in Figures \ref{fig:cp11}(b) and \ref{fig:cp11}(d) show more significant similarity compared with those of spectrograms in Figures \ref{fig:cp11}(a) and \ref{fig:cp11}(c), while they show more significant differences for the same vowel 'iy' spoken by different speakers in Figures \ref{fig:cp11}(b) and \ref{fig:cp11}(f) than that of spectrograms in Figures \ref{fig:cp11}(a) and \ref{fig:cp11}(e). The differences in cochleagram representations for different vowels are more dramatic as in Figures \ref{fig:cp11}(b) and \ref{fig:cp11}(h) than those of spectrograms in Figures \ref{fig:cp11}(a) and \ref{fig:cp11}(g). These all suggest that cochleagrams based on the proposed jointly discretised cochlea model might be better in capturing discriminative information, such as the differences between different vowels and speakers, which may serve as a better front-end extractor for speech-related recognition tasks. 

\section{\label{sec:6} Conclusions}
The front-ends of current speech and audio processing systems are typically time-invariant systems that carry out some forms of spectral analysis of the input speech or audio signals. However, the human cochlea is anything but time-invariant. It continuously adapts to the incoming sound, acting as a spectral analyser with feedback driven dynamic range compression. Numerical models of the cochlea, both with and without this active feedback mechanism have been studied for decades but these models comprise of systems of differential equations and are not amenable for direct use as front-ends in digital speech signal processing systems. In this paper we have derived a joint spatio-temporally discrete model of the cochlea and shown that it can be implemented as a system of coupled difference equations, making it suitable for use in digital systems running at a fixed sampling rate. The proposed model was validated on a range of input signals by comparing with a well established semi-discrete cochlear model. All experimental results indicate that the proposed model matches semi-discrete model in terms of all pertinent characteristics.

The jointly discretised model can be implemented either as a passive model with no feedback elements, or as an active model with linear or non-linear feedback. The parameters used for all three versions are presented. In addition, we present an analyses of the stability of the proposed model with linear feedback and use it to ascertain the resolution required for the temporal discretisation to guarantee stability. Empirical testing of the model implemented with nonlinear feedback at the same temporal resolutions revealed that the nonlinear model was also stable at those resolutions. Finally, we also ascertained that the proposed model is able to achieve around 40dB of dynamic range compression.

Together these characteristics make the proposed jointly discretised model a good model of the human cochlea and a promising biologically-inspired front-end for modern speech processing systems that need to cater to an input signals spanning a large dynamic range. Our future work will focus on implementing the proposed model as layers of a neural network which will allow for integration with a range of state-of-the-art deep learning based speech processing systems, and potentially significantly more complex feedback paths whose parameters can be inferred in a more data driven approach.

\section*{Acknowledgements}
\noindent
This work was funded by Australian Research Council (ARC) Discovery Grant DP190102479.


\section*{APPENDIX A: Semi-Discretised Wave Equation: Spatial Discretisation}\label{app:A} 
\noindent
Typically a finite difference approximation is used to discretise the continuous wave propagation along the spatial dimension, resulting in a system of differential equations (in time) that can be solved numerically. i.e., in the wave equation,

\begin{equation}\tag{A1}
\frac{\partial{p^2(l,t)}}{\partial{l^2}}=\frac{2\rho}{H}\ddot{\xi}(l,t)
\end{equation}

\noindent
approximating the spatial derivative with a finite difference approximation leads to:
\begin{equation}\tag{A2}\label{eq:a2}
\frac{p_{n-1}(t)-2p_n(t)+p_{n+1}(t)}{\Delta l^2}=\frac{2\rho}{H}\ddot{\xi}_n(t)
\end{equation}
\noindent
where $n=2,...,N-1$ represents the $n^{th}$ BM element of the cochlear, with $N$ denoting the total number of cochlear elements in the finite difference approximation, $\ddot{\xi}_{n}$ denoting the transverse acceleration of the $n^{th}$ BM element, and $\Delta l$ the length scale of the spatial discretised BM. At the basal end, $n=0$, and the apical ends, $n=N$, the boundary conditions for the semi-discretised model are given by:
\begin{equation}\tag{A3}\label{eq:a3}
\frac{p_{2}(t)-p_1(t)}{\Delta l}=2\rho \ddot{\xi}_1(t) = 2\rho (\ddot{\xi}_{b}(t) + \ddot{\xi}_{s}(t))
\end{equation}
\begin{equation}\tag{A4}\label{eq:a4}
p_{N}(t)=0
\end{equation}
\noindent
where $\ddot{\xi}_{b}$ represents the acceleration due to the pressure in the ear canal, and $\ddot{\xi}_{s}(t)$ is the acceleration due to the loading by the internal pressure response at the basal end, which is non-zero only for the first BM element.

This system of $N$ equations, corresponding to the $N$ elements of the basilar membrane, forms  the semi-discretised model which can be compactly represented as:
\begin{equation}\tag{A5}\label{eq:a5}
    \mathbf{Fp}(t) - {\mathbf{\ddot\xi}}(t) = \mathbf{q}(t)
\end{equation}
\noindent
where $\boldsymbol{F}$ is the finite-difference matrix:
\begin{equation}\tag{A6}\label{eq:a6}
\mathbf{F} = \frac{H}{2 \rho\Delta l^2}\begin{bmatrix} -\frac{\Delta l}{H}& \frac{\Delta l}{H} & 0 & 0 &  \dots & 0 \\ 1 & -2 & 1 & 0 & \dots & 0 \\ \vdots & \vdots & \vdots & \ddots & \vdots & \vdots \\0  & \dots & 1 & -2 & 1 & 0 \\ 0  & \dots & 0 & 0 & 0  & -\frac{2 \rho\Delta l^2}{H} \end{bmatrix}
\end{equation}
\noindent
$\mathbf{\ddot{\xi}}(t)$ is the vector of BM element accelerations:
\begin{equation}\tag{A7}\mathbf{\ddot{\xi}}(t) = [\ddot{\xi}_1(t),\ddot{\xi}_2(t),\cdots, 0]^T\end{equation}
\noindent
$\mathbf{q}(t)$ with only the first element not zero representing excitation in the form of the source acceleration at the basal end:
\begin{equation}\tag{A8}\mathbf{q}(t) = [\ddot{\xi}_s(t),0,0,\cdots, 0]^T\end{equation}
\noindent
and $\mathbf{p}(t)$ is the $N$ dimensional vector of pressure differences of all elements at time $t$. $\ddot{\mathbf{\xi}}_s(t)$ is the $N$ dimensional vector of source terms where only the first element is non-zero representing the input acceleration at the basal end.

\section*{APPENDIX B: Semi-discretised Micro-mechanical model}\label{app:B}
\noindent
The micro-mechanics of each BM element is described by equations (\ref{eq:5}) and (\ref{eq:6}) which can be written out by taking into account all the forces depicted in Figure \ref{fig:real2}. These two force equations can be rearranged as:
\begin{equation}\tag{B1}\label{eq:b1}
   \ddot{\xi_n^1} = \overline{\alpha}_{2}^n p_n + \overline{\alpha}_{1}^n \dot{\xi_n^1} + \overline{\alpha}_{0}^n \xi_n^1 + \overline{\beta}_{1}^n \dot{\xi_n^2} + \overline{\beta}_{1}^n \xi_n^2  
\end{equation}

\begin{equation}\tag{B2}\label{eq:b2}
    \ddot{\xi_n^2} =  \overline{\delta}_{1}^n \dot{\xi_n^1} + \overline{\delta}_{0}^n \xi_n^1 + \overline{\varepsilon}_{1}^n \dot{\xi_n^2} + \overline{\varepsilon}_{1}^n \xi_n^2
\end{equation}

By choosing to represent the state of each spatial element of the semi-discretised model in terms BM and TM displacements and velocities with the state vector given by $ \mathbf{x}_n(t)=[\dot{\xi_n^1}(t),\xi_n^1(t),\dot{\xi_n^2}(t),\xi_n^2(t)]^T $, the force equations can be compactly represented as:
\begin{equation}\tag{B3}\label{eq:b3}
\begin{split}
\Dot{\mathbf{x}}_n(t)& =
\mathbf{A}_n\mathbf{x}_n(t)+\mathbf{b}_n p(t)   \\
& = \begin{bmatrix} \overline{\alpha}_1^n &  \overline{\alpha}_0^n & \overline{\beta}_1^n & \overline{\beta}_0^n \\ 1 & 0 & 0 & 0 \\ \overline{\delta}_1^n &  \overline{\delta}_0^n & \overline{\varepsilon}_1^n & \overline{\varepsilon}_0^n \\ 0& 0 & 1 & 0 \end{bmatrix} 
\begin{bmatrix}
\dot{\xi_n^1} \\ \xi_n^1 \\ \dot{\xi_n^2} \\ \xi_n^2
\end{bmatrix} + \begin{bmatrix} \overline{\alpha}_{2}^n \\ 0 \\ 0  \\ 0  \end{bmatrix} p_n
\end{split}
\end{equation}
\noindent
where the elements of $\mathbf{A}_n$ and $\mathbf{b}_n$ are given as follows:
\begin{equation}\tag{B4}
\overline{\alpha}_1^n = -\frac{c_1+gc_3-\gamma gc_4}{m_1}
\end{equation}

\begin{equation}\tag{B5}
\overline{\alpha}_0^n =-\frac{k_1+gk_3-\gamma gc_4}{m_1}
\end{equation}

\begin{equation}\tag{B6}
\overline{\beta}_{1}^n = \frac{c_3-\gamma c_4}{m_1}
\end{equation}

\begin{equation}\tag{B7}
\overline{\beta}_{0}^n = \frac{k_3-\gamma k_4}{m_1}
\end{equation}

\begin{equation}\tag{B8}
\overline{\delta}_1^n =\frac{c_3}{m_2}
\end{equation}

\begin{equation}\tag{B9}
\overline{\delta}_0^n =\frac{k_3}{m_2}
\end{equation}

\begin{equation}\tag{B10}
\overline{\varepsilon}_1^n = -\frac{c_2+c_3}{m_2}
\end{equation}

\begin{equation}\tag{B11}
\overline{\varepsilon}_0^n = -\frac{k_2 + k_3}{m_2}
\end{equation}

\noindent
Extending to all $N$ spatial elements, the micro-mechanical model can be written as:

\begin{equation}\tag{B12}\label{eq:b12}
    \Dot{\mathbf{x}}(t)=\mathbf{A}_E\mathbf{x}(t)+\mathbf{B}_E\mathbf{p}(t)
\end{equation}

\noindent
where $\mathbf{A}_E$ is the $4N \times 4N$ block diagonal matrix:
\begin{equation}\tag{B13}
     \mathbf{A}_E = \begin{bmatrix}
   \mathbf{A}_1 & 0 & 0 & \cdots & 0 & 0 \\ 0 & \mathbf{A}_2 & 0 & \cdots & 0 & 0  \\
   \vdots & \vdots & \vdots & \vdots & \ddots & \vdots  \\
   0 & 0 & 0 & \cdots & \mathbf{A}_{N-1} & 0 \\0 & 0 & 0 & 0 & \cdots & \mathbf{A}_N
   \end{bmatrix}
\end{equation}
\noindent
and $\mathbf{B}_E$ is the $4N \times N$ matrix:
\begin{equation}\tag{B14}
     \mathbf{B}_E = \begin{bmatrix}
   \mathbf{b}_1 & 0 & 0 & \cdots & 0 & 0 \\ 0 & \mathbf{b}_2 & 0 & \cdots & 0 & 0  \\
   \vdots & \vdots & \vdots & \vdots & \ddots & \vdots  \\
   0 & 0 & 0 & \cdots & \mathbf{b}_{N-1} & 0 \\0 & 0 & 0 & 0 & \cdots & \mathbf{b}_N
   \end{bmatrix}
\end{equation}
\noindent
with the diagonal elements $\mathbf{A}_n$ and $\mathbf{b}_n$ given as per equation (\ref{eq:b3}).

\section*{APPENDIX C: Semi-Discretised State Space Model: State Space Formulation}\label{app:C}
\noindent
The state space formulation of the semi-discretised model is obtained by observing that:
\begin{equation}\tag{C1}\label{eq:c1}
    \mathbf{\dot{\xi}}(t)=\mathbf{C}_E\mathbf{x}(t)
\end{equation}

\noindent
where $\mathbf{C}_E$ is the output selection matrix which selects the variable of interest from the state vector (such as BM velocities) and is given by:
\begin{equation}\tag{C2}\label{eq:c2}
     \mathbf{C}_E = \begin{bmatrix}
   1 & 0 & 0 & 0 & \cdots & \cdots & \cdots & \cdots &  0 \\ 0 & 0 & 0 & 0 & 1 & 0 & \cdots & \cdots & 0  \\
   \vdots & \vdots  & \vdots & \vdots & \vdots & \vdots & \vdots & \ddots & \vdots \\0 & 0 & \cdots & \cdots & \cdots & 1 & 0 & 0 & 0
   \end{bmatrix}
\end{equation}
This allows the wave equation (\ref{eq:a5}) to be written as:
\begin{equation}\tag{C3}\label{eq:c3}
    \mathbf{Fp}(t) - \mathbf{C   }_{E}{\mathbf{\dot{x}}(t)} = \mathbf{q}(t)
\end{equation}
which, when combined with the micro-mechanical model (\ref{eq:b12}), leads to the state space representation:
\begin{equation}\tag{C4}\label{eq:c4}
    \mathbf{\dot{x}}(t)=\mathbf{A}\mathbf{x}(t)+\mathbf{B}\mathbf{u}(t)
\end{equation}
\noindent
where system matrices $\mathbf{A}$ and $\mathbf{B}$ are:
\begin{equation}\tag{C5}\label{eq:c5}
    \mathbf{A}=\mathbf{(I-B}_E\mathbf{F}^{-1}\mathbf{C}_E)^{-1}\mathbf{A}_E
\end{equation}
\begin{equation}\tag{C6}\label{eq:c6}
    \mathbf{B}=\mathbf{(I-B}_E\mathbf{F}^{-1}\mathbf{C}_E)^{-1}\mathbf{B}_E
\end{equation}
and $\mathbf{u}(t)$ is the input stimulus with only the first element being non-zero \cite{elliott2007state}.

\section*{APPENDIX D: Joint spatio-temporal discretised model}\label{app:D}
\noindent
Summing all the forces action on the BM (see Figure \ref{fig:real2}) leads to the following equation:
\begin{equation}\tag{D1}\label{eq:d1}
\begin{split}
    p(l,t) - p_a(l,t) &= m_1\ddot{\xi^1}(l,t)+c_1(l)\dot{\xi^1}(l,t)
    \\ & +k^1(l)\xi^1(l,t) \\ &+c_3(l)\dot{\xi^f}(l,t)+k_3(l)\xi^f(l,t)
\end{split}
\end{equation}
\noindent
where, $p(l,t)$ denotes the pressure difference between the upper and lower chambers of the cochlea at position $l$; $p_a(l,t)$ denotes the additional active feedback pressure generated by OHCs; $m_1$ denotes the BM segment mass; $c_1(l)$ and $k_1(l)$ represent BM damper and stiffness coefficients; $c_3(l)$ and $k_3(l)$ represent the damper and stiffness coefficient corresponding to the fluid coupling between TM and BM; and $\xi^f(l,t)$ represents the relative displacement between TM and RL, where the displacement of RL is proportional to BM displacement with $g$ as the constant of proportionality:
\begin{equation}\tag{D2}\label{eq:d2}
    \xi^f(l,t)=g\xi^1(l,t)-\xi^2(l,t)
\end{equation}
\noindent
Similarly, the forces acting on the TM segment with mass, $m_2$, lead to:
\begin{equation}\tag{D3}\label{eq:d3}
\begin{split}
    m_2\ddot{\xi^2}(l,t)+&c_2(l)\dot{\xi^2}(l,t)+k_2(l)\xi^2(l,t) \\
    &= c_3(l)\dot{\xi^f}(l,t)+k_3(l)\xi^f(l,t)
\end{split}
\end{equation}


\noindent
Discretising equations (\ref{eq:d1}) and (\ref{eq:d3}), both spatially and temporally, and combining with (\ref{eq:d2}) leads to:

\begin{equation}\tag{D4}\label{eq:d4}
\begin{split}
  p_{j,n} &+ \gamma c_4 (g \frac{\xi_{j+1,n}^{1}-\xi_{j,n}^1}{\Delta_{t}} - \frac{\xi_{j+1,n}^{2}-\xi_{j,n}^2}{\Delta_{t}})  \\
  & +k_4(g\xi_{j,n}^{1}-\xi_{j,n}^2)  \\
  &=m_1\frac{\xi_{j+1,n}^1-2\xi_{j,n}^1+\xi_{j-1,n}^1}{\Delta_{t}^2}\\
 & +c_1 \frac{\xi_{j+1,n}^{1}-\xi_{j,n}^1}{\Delta_{t}}+k_1\xi_{j,n}^1 \\
 & + c_3 (g \frac{\xi_{j+1,n}^{1}-\xi_{j,n}^1}{\Delta_{t}} - \frac{\xi_{j+1,n}^{2}-\xi_{j,n}^2}{\Delta_{t}})  \\
&  + k_3(\xi_{j,n}^{1}-\xi_{j,n}^2)
\end{split}
\end{equation}
and,
\begin{equation}\tag{D5}\label{eq:d5}
\begin{split}
 & m_2 \frac{\xi_{j+1,n}^{2}-2\xi_{j,n}^2+\xi_{j-1,n}^{2}}{\Delta_{t}^2} +
 c_2 \frac{\xi_{j+1,n}^{2}-\xi_{j,n}^2}{\Delta_{t}} \\  & +k_2\xi_{j,n}^2 \\  
 & = c_3 (g*\frac{\xi_{j+1,n}^{1}-\xi_{j,n}^1}{\Delta_{t}} - \frac{\xi_{j+1,n}^{2}-\xi_{j,n}^2}{\Delta_{t}})  \\
&  + k_3(g\xi_{j,n}^{1}-\xi_{j,n}^2)
\end{split}
\end{equation}

Please note that the mechanical parameters of the BM and TM vary with position but we have simplified the notation for ease of readability and do not explicitly denote the dependence on $n$ for the parameters $c_1, c_2, c_3, c_4, k_1, k_2, k_3$, and $k_4$. Rearranging the terms in both force equations lead to (\ref{eq:18}) and (\ref{eq:19}), which are both repeated here:
\begin{equation}\tag{D6}
\begin{split}
  p_{j,n} = & \alpha_{n}^1 \xi_{j+1,n}^{1} + \alpha_{n}^0 \xi_{j,n}^{1} + \alpha_{n}^{-1}\xi_{j-1,n}^{1}  \\
  & +\beta_{n}^1 \xi_{j+1,n}^{2} + \beta_{n}^0 \xi_{j,n}^{2}
\end{split}
\end{equation}
and,
\begin{equation}\tag{D7}
\begin{split}
  0 = & \varepsilon_n^1 \xi_{j+1,n}^{1} + \varepsilon_n^0 \xi_{j,n}^{1} +\delta_n^1 \xi_{j+1,n}^{2} \\
  &  + \delta_n^0 \xi_{j,n}^{2} + \delta_n^{-1} \xi_{j-1,n}^{2}
\end{split}  
\end{equation}

\noindent
The coefficients in these two equations are given below:
\begin{equation}\tag{D8}
\alpha_{n}^1 = \frac{m_1}{\Delta_{t}^2} +\frac{c_1+gc_3-\gamma gc_4}{\Delta_{t}}
\end{equation}

\begin{equation}\tag{D9}
\alpha_{n}^0 = -\frac{2m_1}{\Delta_{t}^2}-\frac{c_1+gc_3-\gamma gc_4}{\Delta_{t}} +(k_1+gk_3-\gamma gk_4)
\end{equation}

\begin{equation}\tag{D10}
\alpha_{n}^{-1} = \frac{m_1}{\Delta_{t}^2}
\end{equation}

\begin{equation}\tag{D11}
\beta_{n}^{1} = \frac{\gamma c_4-c_3}{\Delta_{t}}
\end{equation}

\begin{equation}\tag{D12}
\beta_{n}^{0} = -\frac{\gamma c_4-c_3}{\Delta_{t}}+(\gamma k_4-k_3)
\end{equation}

\begin{equation}\tag{D13}
\delta_{n}^1 = -\frac{m_2}{\Delta_{t}^2} -\frac{c_2+c_3}{\Delta_{t}}
\end{equation}

\begin{equation}\tag{D14}
\delta_{n}^0 = \frac{2m_2}{\Delta_{t}^2} +\frac{c_2+c_3}{\Delta_{t}} - (k_2+k_3)
\end{equation}

\begin{equation}\tag{D15}
\delta_{n}^{-1} = -\frac{m_2}{\Delta_{t}^2}
\end{equation}

\begin{equation}\tag{D16}
\varepsilon_{n}^{1} = \frac{gc_3}{\Delta_{t}}
\end{equation}

\begin{equation}\tag{D17}
\varepsilon_{n}^{0} = -\frac{gc_3}{\Delta_{t}}+k_3g
\end{equation}

\section*{APPENDIX E: Parameters in passive model}\label{app:E}
\begin{table}[ht!]\label{tab:1}
    \caption{Parameters of the passive jointly discretised cochlear model.}
    \centering
    \begin{tabular}{c|c}
    \hline
    Parameter & Value (SI)\\
    \hline
     $m_1$ & 0.28 $kgm^{-2}$\\
     Q & 5 \\
     $k_1$ & 2$\pi f^2$ $m_1Nm^{-3}$ \\
     $c_1$ & $\frac{\sqrt{k_1m_1}}{Q}Nsm^{-3}$\\
     $m_{ME}$ & 1.4080 $kgm^{-2}$\\
     $k^{ME}$ & 2.592*$10^8$ $Nm^{-3}$\\
     $c^{ME}$ & 32000 $Nsm^{-3}$\\
     \hline
    \end{tabular}
\end{table}

\bibliography{sampbib}

\begin{thebibliography}{10}
\def\enquote#1,{``#1,''}
\expandafter\ifx\csname url\endcsname\relax
  \def\url#1{\texttt{#1}}\fi
\expandafter\ifx\csname urlprefix\endcsname\relax\def\urlprefix{URL }\fi
\providecommand{\bibinfo}[2]{#2}
\def\plainquote#1{``#1''}
\providecommand{\noopsort}[1]{}
\providecommand{\switchargs}[2]{#2#1}
\providecommand{\dourl}[1]{\href{http://#1}{\nolinkurl{#1}}}
\providecommand{\dodoi}[1]{doi: \href{http://dx.doi.org/#1}{\nolinkurl{#1}}}
  \def\eatspace #1{#1}

\bibitem{elliott2012cochlea}
\bibinfo{author}{S.~J. Elliott} and \bibinfo{author}{C.~A. Shera},
  \enquote{\bibinfo{title}{The cochlea as a smart structure}},
  \bibinfo{journal}{Smart Materials and Structures} \textbf{21}(6),
  \bibinfo{pages}{064001} (\bibinfo{year}{2012}).

\bibitem{camalet2000auditory}
\bibinfo{author}{S.~Camalet}, \bibinfo{author}{T.~Duke},
  \bibinfo{author}{F.~J{\"u}licher}, and \bibinfo{author}{J.~Prost},
  \enquote{\bibinfo{title}{Auditory sensitivity provided by self-tuned critical
  oscillations of hair cells}}, \bibinfo{journal}{Proceedings of the national
  academy of sciences} \textbf{97}(7), \bibinfo{pages}{3183--3188}
  (\bibinfo{year}{2000}).

\bibitem{von1960experiments}
\bibinfo{author}{G.~Von~B{\'e}k{\'e}sy} and \bibinfo{author}{E.~G. Wever},
  \emph{\bibinfo{title}{Experiments in hearing}}, Vol.~\bibinfo{volume}{8}
  (\bibinfo{publisher}{McGraw-Hill New York}, \bibinfo{year}{1960}).

\bibitem{moore2012introduction}
\bibinfo{author}{B.~C. Moore}, \emph{\bibinfo{title}{An introduction to the
  psychology of hearing}}  (\bibinfo{publisher}{Brill}, \bibinfo{year}{2012}).

\bibitem{steele1974behavior}
\bibinfo{author}{C.~Steele}, \enquote{\bibinfo{title}{Behavior of the basilar
  membrane with pure-tone excitation}}, \bibinfo{journal}{The Journal of the
  Acoustical Society of America} \textbf{55}(1), \bibinfo{pages}{148--162}
  (\bibinfo{year}{1974}).

\bibitem{steele1980improved}
\bibinfo{author}{C.~R. Steele} and \bibinfo{author}{C.~E. Miller},
  \enquote{\bibinfo{title}{An improved wkb calculation for a two-dimensional
  cochlear model}}, \bibinfo{journal}{The Journal of the Acoustical Society of
  America} \textbf{68}(1), \bibinfo{pages}{147--148} (\bibinfo{year}{1980}).

\bibitem{taber1981cochlear}
\bibinfo{author}{L.~A. Taber} and \bibinfo{author}{C.~R. Steele},
  \enquote{\bibinfo{title}{Cochlear model including three-dimensional fluid and
  four modes of partition flexibility}}, \bibinfo{journal}{The Journal of the
  Acoustical Society of America} \textbf{70}(2), \bibinfo{pages}{426--436}
  (\bibinfo{year}{1981}).

\bibitem{lim2002three}
\bibinfo{author}{K.-M. Lim} and \bibinfo{author}{C.~R. Steele},
  \enquote{\bibinfo{title}{A three-dimensional nonlinear active cochlear model
  analyzed by the wkb-numeric method}}, \bibinfo{journal}{Hearing research}
  \textbf{170}(1-2), \bibinfo{pages}{190--205} (\bibinfo{year}{2002}).

\bibitem{kanis1993self}
\bibinfo{author}{L.~J. Kanis} and \bibinfo{author}{E.~de~Boer},
  \enquote{\bibinfo{title}{Self-suppression in a locally active nonlinear model
  of the cochlea: A quasilinear approach}}, \bibinfo{journal}{The Journal of
  the Acoustical Society of America} \textbf{94}(6),
  \bibinfo{pages}{3199--3206} (\bibinfo{year}{1993}).

\bibitem{chadwick1998compression}
\bibinfo{author}{R.~Chadwick}, \enquote{\bibinfo{title}{Compression, gain, and
  nonlinear distortion in an active cochlear model with subpartitions}},
  \bibinfo{journal}{Proceedings of the National Academy of Sciences}
  \textbf{95}(25), \bibinfo{pages}{14594--14599} (\bibinfo{year}{1998}).

\bibitem{ni2014modelling}
\bibinfo{author}{G.~Ni}, \bibinfo{author}{S.~J. Elliott},
  \bibinfo{author}{M.~Ayat}, and \bibinfo{author}{P.~D. Teal},
  \enquote{\bibinfo{title}{Modelling cochlear mechanics}},
  \bibinfo{journal}{BioMed research international} \textbf{2014}
  (\bibinfo{year}{2014}).

\bibitem{elliott2007state}
\bibinfo{author}{S.~J. Elliott}, \bibinfo{author}{E.~M. Ku}, and
  \bibinfo{author}{B.~Lineton}, \enquote{\bibinfo{title}{A state space model
  for cochlear mechanics}}, \bibinfo{journal}{The Journal of the Acoustical
  Society of America} \textbf{122}(5), \bibinfo{pages}{2759--2771}
  (\bibinfo{year}{2007}).

\bibitem{sharan2015cochleagram}
\bibinfo{author}{R.~V. Sharan} and \bibinfo{author}{T.~J. Moir},
  \enquote{\bibinfo{title}{Cochleagram image feature for improved robustness in
  sound recognition}}, in \emph{\bibinfo{booktitle}{2015 IEEE International
  Conference on Digital Signal Processing (DSP)}}, \bibinfo{organization}{IEEE}
  (\bibinfo{year}{2015}), pp. \bibinfo{pages}{441--444}.

\bibitem{buermannspeech}
\bibinfo{author}{M.~Buermann} and \bibinfo{author}{T.~A. van Meer},
  \enquote{\bibinfo{title}{Speech recognition using very deep neural networks:
  Spectrograms vs cochleagrams}},   (\bibinfo{year}{2020}).

\bibitem{koizumi1996speech}
\bibinfo{author}{T.~Koizumi}, \bibinfo{author}{M.~Mori}, and
  \bibinfo{author}{S.~Taniguchi}, \enquote{\bibinfo{title}{Speech recognition
  based on a model of human auditory system}}, in
  \emph{\bibinfo{booktitle}{Proceeding of Fourth International Conference on
  Spoken Language Processing. ICSLP'96}}, \bibinfo{organization}{IEEE}
  (\bibinfo{year}{1996}), Vol.~\bibinfo{volume}{2}, pp.
  \bibinfo{pages}{937--940}.

\bibitem{ting2004speaker}
\bibinfo{author}{H.~N. Ting} and \bibinfo{author}{J.~Yunus},
  \enquote{\bibinfo{title}{Speaker-independent malay vowel recognition of
  children using multi-layer perceptron}}, in \emph{\bibinfo{booktitle}{2004
  IEEE Region 10 Conference TENCON 2004.}}, \bibinfo{organization}{IEEE}
  (\bibinfo{year}{2004}), pp. \bibinfo{pages}{68--71}.

\bibitem{lyon2011cascades}
\bibinfo{author}{R.~F. Lyon}, \enquote{\bibinfo{title}{Cascades of
  two-pole--two-zero asymmetric resonators are good models of peripheral
  auditory function}}, \bibinfo{journal}{The Journal of the Acoustical Society
  of America} \textbf{130}(6), \bibinfo{pages}{3893--3904}
  (\bibinfo{year}{2011}).

\bibitem{baby2020convolutional}
\bibinfo{author}{D.~Baby}, \bibinfo{author}{A.~V.~D. Broucke}, and
  \bibinfo{author}{S.~Verhulst}, \enquote{\bibinfo{title}{A convolutional
  neural-network model of human cochlear mechanics and filter tuning for
  real-time applications}}, \bibinfo{journal}{arXiv preprint arXiv:2004.14832}
  (\bibinfo{year}{2020}).

\bibitem{neely1986model}
\bibinfo{author}{S.~T. Neely} and \bibinfo{author}{D.~Kim},
  \enquote{\bibinfo{title}{A model for active elements in cochlear
  biomechanics}}, \bibinfo{journal}{The journal of the acoustical society of
  America} \textbf{79}(5), \bibinfo{pages}{1472--1480} (\bibinfo{year}{1986}).

\bibitem{de1996mechanics}
\bibinfo{author}{E.~De~Boer}, \enquote{\bibinfo{title}{Mechanics of the
  cochlea: modeling efforts}}, in \emph{\bibinfo{booktitle}{The cochlea}}
  (\bibinfo{publisher}{Springer}, \bibinfo{year}{1996}), pp.
  \bibinfo{pages}{258--317}.

\bibitem{diependaal1989time}
\bibinfo{author}{R.~J. Diependaal}, \enquote{\bibinfo{title}{Time-domain
  solutions for 1d, 2d and 3d cochlear models}}, in
  \emph{\bibinfo{booktitle}{Cochlear Mechanisms: Structure, Function, and
  Models}}  (\bibinfo{publisher}{Springer}, \bibinfo{year}{1989}), pp.
  \bibinfo{pages}{445--452}.

\bibitem{steele1979comparison}
\bibinfo{author}{C.~R. Steele} and \bibinfo{author}{L.~A. Taber},
  \enquote{\bibinfo{title}{Comparison of wkb and finite difference calculations
  for a two-dimensional cochlear model}}, \bibinfo{journal}{The Journal of the
  Acoustical Society of America} \textbf{65}(4), \bibinfo{pages}{1001--1006}
  (\bibinfo{year}{1979}).

\bibitem{diependaal1989nonlinear}
\bibinfo{author}{R.~J. Diependaal} and \bibinfo{author}{M.~A. Viergever},
  \enquote{\bibinfo{title}{Nonlinear and active two-dimensional cochlear
  models: Time-domain solution}}, \bibinfo{journal}{The Journal of the
  Acoustical Society of America} \textbf{85}(2), \bibinfo{pages}{803--812}
  (\bibinfo{year}{1989}).

\bibitem{neely1981finite}
\bibinfo{author}{S.~T. Neely}, \enquote{\bibinfo{title}{Finite difference
  solution of a two-dimensional mathematical model of the cochlea}},
  \bibinfo{journal}{The Journal of the Acoustical Society of America}
  \textbf{69}(5), \bibinfo{pages}{1386--1393} (\bibinfo{year}{1981}).

\bibitem{steele1979comparison1}
\bibinfo{author}{C.~R. Steele} and \bibinfo{author}{L.~A. Taber},
  \enquote{\bibinfo{title}{Comparison of wkb calculations and experimental
  results for three-dimensional cochlear models}}, \bibinfo{journal}{The
  Journal of the Acoustical Society of America} \textbf{65}(4),
  \bibinfo{pages}{1007--1018} (\bibinfo{year}{1979}).

\bibitem{elliott2018elemental}
\bibinfo{author}{S.~J. Elliott} and \bibinfo{author}{G.~Ni},
  \enquote{\bibinfo{title}{An elemental approach to modelling the mechanics of
  the cochlea}}, \bibinfo{journal}{Hearing research} \textbf{360},
  \bibinfo{pages}{14--24} (\bibinfo{year}{2018}).

\bibitem{rhode1971observations}
\bibinfo{author}{W.~S. Rhode}, \enquote{\bibinfo{title}{Observations of the
  vibration of the basilar membrane in squirrel monkeys using the m{\"o}ssbauer
  technique}}, \bibinfo{journal}{The Journal of the Acoustical Society of
  America} \textbf{49}(4B), \bibinfo{pages}{1218--1231} (\bibinfo{year}{1971}).

\bibitem{robles1976transient}
\bibinfo{author}{L.~Robles}, \bibinfo{author}{W.~S. Rhode}, and
  \bibinfo{author}{C.~D. Geisler}, \enquote{\bibinfo{title}{Transient response
  of the basilar membrane measured in squirrel monkeys using the m{\"o}ssbauer
  effect}}, \bibinfo{journal}{The Journal of the Acoustical Society of America}
  \textbf{59}(4), \bibinfo{pages}{926--939} (\bibinfo{year}{1976}).

\bibitem{johnstone1986basilar}
\bibinfo{author}{B.~Johnstone}, \bibinfo{author}{R.~Patuzzi}, and
  \bibinfo{author}{G.~Yates}, \enquote{\bibinfo{title}{Basilar membrane
  measurements and the travelling wave}}, \bibinfo{journal}{Hearing research}
  \textbf{22}(1-3), \bibinfo{pages}{147--153} (\bibinfo{year}{1986}).

\bibitem{ruggero1991furosemide}
\bibinfo{author}{M.~A. Ruggero} and \bibinfo{author}{N.~C. Rich},
  \enquote{\bibinfo{title}{Furosemide alters organ of corti mechanics: evidence
  for feedback of outer hair cells upon the basilar membrane}},
  \bibinfo{journal}{Journal of Neuroscience} \textbf{11}(4),
  \bibinfo{pages}{1057--1067} (\bibinfo{year}{1991}).

\bibitem{ruggero1992responses}
\bibinfo{author}{M.~A. Ruggero}, \enquote{\bibinfo{title}{Responses to sound of
  the basilar membrane of the mammalian cochlea}}, \bibinfo{journal}{Current
  opinion in neurobiology} \textbf{2}(4), \bibinfo{pages}{449--456}
  (\bibinfo{year}{1992}).

\bibitem{lyon1988cochlear}
\bibinfo{author}{R.~F. Lyon} and \bibinfo{author}{C.~A. Mead},
  \enquote{\bibinfo{title}{Cochlear hydrodynamics demystified}},
  (\bibinfo{year}{1988}).

\bibitem{lyon1990automatic}
\bibinfo{author}{R.~F. Lyon}, \enquote{\bibinfo{title}{Automatic gain control
  in cochlear mechanics}}, in \emph{\bibinfo{booktitle}{The mechanics and
  biophysics of hearing}}  (\bibinfo{publisher}{Springer},
  \bibinfo{year}{1990}), pp. \bibinfo{pages}{395--402}.

\bibitem{neely1983active}
\bibinfo{author}{S.~T. Neely} and \bibinfo{author}{D.~O. Kim},
  \enquote{\bibinfo{title}{An active cochlear model showing sharp tuning and
  high sensitivity}}, \bibinfo{journal}{Hearing research} \textbf{9}(2),
  \bibinfo{pages}{123--130} (\bibinfo{year}{1983}).

\bibitem{wilson1980evidence}
\bibinfo{author}{J.~Wilson}, \enquote{\bibinfo{title}{Evidence for a cochlear
  origin for acoustic re-emissions, threshold fine-structure and tonal
  tinnitus}}, \bibinfo{journal}{Hearing research} \textbf{2}(3-4),
  \bibinfo{pages}{233--252} (\bibinfo{year}{1980}).

\bibitem{allen2001nonlinear}
\bibinfo{author}{J.~Allen}, \enquote{\bibinfo{title}{Nonlinear cochlear signal
  processing}}, in \emph{\bibinfo{booktitle}{Physiology of the Ear, Second
  Edition}}  (\bibinfo{publisher}{Singular Thompson}, \bibinfo{year}{2001}),
  pp. \bibinfo{pages}{393--442}.

\bibitem{ku2008modelling}
\bibinfo{author}{E.~M. Ku}, \enquote{\bibinfo{title}{Modelling the human
  cochlea}}, Ph.D. thesis, \bibinfo{school}{University of Southampton},
  \bibinfo{year}{2008}.

\bibitem{ku2008statistics}
\bibinfo{author}{E.~M. Ku}, \bibinfo{author}{S.~J. Elliott}, and
  \bibinfo{author}{B.~Lineton}, \enquote{\bibinfo{title}{Statistics of
  instabilities in a state space model of the human cochlea}},
  \bibinfo{journal}{The Journal of the Acoustical Society of America}
  \textbf{124}(2), \bibinfo{pages}{1068--1079} (\bibinfo{year}{2008}).

\bibitem{terhardt1979calculating}
\bibinfo{author}{E.~Terhardt}, \enquote{\bibinfo{title}{Calculating virtual
  pitch}}, \bibinfo{journal}{Hearing research} \textbf{1}(2),
  \bibinfo{pages}{155--182} (\bibinfo{year}{1979}).

\bibitem{zue1990speech}
\bibinfo{author}{V.~Zue}, \bibinfo{author}{S.~Seneff}, and
  \bibinfo{author}{J.~Glass}, \enquote{\bibinfo{title}{Speech database
  development at mit: Timit and beyond}}, \bibinfo{journal}{Speech
  communication} \textbf{9}(4), \bibinfo{pages}{351--356}
  (\bibinfo{year}{1990}).

\end{thebibliography}
\end{document}